\documentclass{JHEP3} 

\usepackage{epsfig,multicol}

\newcommand{\no}{\nonumber\\}
\newcommand{\be}{\begin{equation}}
\newcommand{\ee}{\end{equation}}
\newcommand{\ba}{\begin{eqnarray}}
\newcommand{\ea}{\end{eqnarray}}
\newcommand{\ci}[1]{\cite{#1}}
\newcommand{\bi}[1]{\bibitem{#1}}
\newcommand{\la}[1]{\label{#1}}
\def\gl#1{(\ref{#1})}

\newcommand\fverb{\setbox\pippobox=\hbox\bgroup\verb}
\newcommand\fverbdo{\egroup\medskip\noindent%
                        \fbox{\unhbox\pippobox}\ }
\newcommand\fverbit{\egroup\item[\fbox{\unhbox\pippobox}]}
\newbox\pippobox

\title{ Domain wall generation by fermion self-interaction and
light particles}

\author{A. A. Andrianov $^{\flat\sharp}$, 
V.A.Andrianov $^{\flat}$, 
P. Giacconi $^{\sharp}$, 
R. 
Soldati$^{\sharp}$\\
$^{\flat}$ V.A.Fock Department of Theoretical Physics,
Sankt-Petersburg State University,\\
198504 Sankt-Petersburg, Russia\\
$^{\sharp}$ Dipartimento di Fisica, 
Universit\'a
di Bologna and\\
 Istituto Nazionale di Fisica Nucleare, Sezione di 
Bologna,\\
40126 Bologna, Italia}

\abstract{A possible explanation for the appearance of light fermions and  
Higgs bosons on the
four-dimensional domain wall is proposed.
The mechanism of light particle trapping
is accounted for by a strong self-interaction of five-dimensional pre-quarks.
We obtain the low-energy effective action which exhibits the
invariance under the so called $\tau$-symmetry. Then
we find a set of vacuum solutions which break that symmetry and the
five-dimensional translational invariance. 
One type of those vacuum solutions gives rise to the domain wall formation 
with consequent
trapping of light massive fermions and Higgs-like bosons as well as
massless sterile scalars, 
the so-called branons. The induced relations between low-energy 
couplings for
Yukawa and scalar field interactions allow to
make certain predictions for light particle masses and couplings themselves,
which might provide a signature of the
higher dimensional origin of particle physics at future experiments.
The manifest translational symmetry breaking, eventually due to some
gravitational and/or matter fields in five dimensions, is
effectively realized with the help of background scalar defects. As a result
the branons acquire masses, whereas the ratio of Higgs and fermion
(presumably top-quark)
masses can be reduced towards the values compatible with the present-day
phenomenology. Since the branons do not couple to fermions and the Higgs bosons
do not decay into branons, the latter ones are essentially sterile and stable, 
what makes them the natural candidates for the dark matter in the Universe.}
 
\keywords{eld.ssb.bsm}

\begin{document}

\section{Introduction}
The accommodation of our matter world on a four-dimensional surface --
a domain wall or a thick 3-brane -- in
a multi-dimensional space-time with dimension higher than four has
recently attracted much interest as a theoretical concept 
%\ci{R-S,Akama,ADD,AADD,Gog,RS1} 
\ci{R-S}-\ci{RS1} 
promoting novel solutions to the long standing problems of 
the Planck mass scale \ci{ADD,AADD},
GUT scale \ci{gut1,gut2},
SUSY breaking scale \ci{susyb1,susyb2}, 
electroweak breaking scale 
%\ci{ewb1,ewb2,ewb3,ewb4}
\ci{ewb1}-\ci{ewb4}
and fermion mass hierarchy \ci{fmasshi1,fmasshi2}.
Respectively, an experimental challenge has been posed for the forthcoming
collider and non-collider physics programs to discover new particles, such as
 Kaluza-Klein gravitons
\ci{KKgrav1,KKgrav2},
radions and graviscalars \ci{rad1,rad2}, branons
%\ci{bran1,bran2,bran3},
\ci{bran1}-\ci{bran3}, 
sterile neutrinos 
\ci{gut1,sneut1,sneut2,moha} etc., together with some other
missing energy \ci{misse} or missing charge effects \ci{missch}.
The vast literature on those subjects
and their applications is now covered in few review articles
%\ci{rev1,rev2,rev3,rev4,rev5}.
\ci{rev1}-\ci{rev5}.
Typically, the thick or fat domain wall formation and the trapping
of low-energy particles
on its surface (layer) might be obtained 
%\ci{galoc1,galoc2,galoc3} 
\ci{galoc1}-\ci{galoc3} by a number of particular background
scalar and/or gravitational fields living in the multi-dimensional bulk --
see however Ref.~\ci{vol} --
the configuration of which does generate
zero-energy states localized on the brane.
Obviously, the explanation of how such background fields can emerge and induce
the spontaneous symmetry breaking is to be found
and the  domain wall creation, due to self-interaction
of certain particles in the bulk with low-energy counterparts
populating our world, may be one conceivable possibility.

In this paper we consider and explore the non-compact 
{\it 4 + 1}-dimensional fermion model
with strong local four-fermion interaction that leads to the discrete symmetry
breaking, owing to which the domain wall pattern of the vacuum state
just arises and allows the light massive Dirac particles to live
in four dimensions\footnote{ One can find some relationship of this mechanism
of domain-wall generation to that one of the Top-Mode Standard Model
 \ci{Topmode} used to supply the top-quark with a large mass 
in four dimensions due to quark condensation uniform in the space-time. 
However, in our case, the vacuum state will receive
a scalar background defect breaking translational invariance. 
As well, our model
is taken five-dimensional and non-compact as compared to  six- 
or eight-dimensional extensions of the Top-Mode 
Standard Model \ci{6top1,6top2} with compact extra dimensions and an
essential role played by Kaluza-Klein gravitons and/or gauge fields.}.  In
such a 
model the  scalar fields appear to be as composite ones. We are aware of the
possible important role \ci{rev2,galoc2} 
of a non-trivial gravitational background, of
 propagating gravitons and gauge fields  for issues of stability 
of the domain wall induced by a 
fermion self-interaction. Nonetheless, in order to keep track
of the main dynamical mechanism, we simplify
herein the fermion model just neglecting all gravitational and gauge field
interaction and, in this sense, our model might be thought as a sector
of the full Domain-Wall
Standard Model. However, we believe that the simplified
model we treat in the present investigation will be able to
give us the plenty of quantitative
hints on the relationships among physical characteristics of light particles
trapped on the brane.

Let us elucidate the domain wall phenomenology
in more details and start from the model of
one five-dimensional fermion bi-spinor $\psi(X)$
coupled to a scalar field $\Phi(X)$.
The extra-dimension coordinate is assumed to be space-like,
$$
(X_\alpha) = (x_\mu, z)\ , \quad (x_\mu) = (x_0, x_1, x_2, x_3)\ , \quad
(g_{\alpha\alpha}) = (+,-,-,-,-)
$$ 
and the subspace of $x_\mu$ eventually corresponds to the 
four-dimensional Minkowski
space. The extra-dimension size is supposed to be infinite (or large enough).
The fermion wave function is then described by the Dirac equation
\be
[\,i\gamma_\alpha \partial^\alpha - \Phi(X)\,]\psi(X) = 0\ , \quad
\gamma_\alpha = (\gamma_\mu, -i\gamma_5)\ ,\quad \{\gamma_\alpha, 
\gamma_\beta\}
= 2g_{\alpha\beta}\ , \la{5dir}
\ee
with $\gamma_\alpha$ being a set of four-dimensional Dirac matrices in the 
chiral representation. 

The trapping of 
light fermions on a four-dimensional hyper-plane  -- the domain wall --
localized in the
fifth dimension at $z = z_0$ can be promoted
by a certain {\sl topological}, $z$-dependent background configuration of
the scalar field  $\langle\Phi(X)\rangle_0 = \varphi (z)$,
due to the appearance of zero-modes in the four-dimensional fermion spectrum.
In this case, from the viewpoint of four-dimensional space-time,
Eq.\,\gl{5dir} just characterizes the infinite set of fermions
with different masses
that it is easier to see after the Fock-Schwinger transformation,
\ba
&&[\,i\gamma_\alpha \partial^\alpha + \varphi(z)\,][\,i\gamma_\alpha
\partial^\alpha - \varphi(z)\,]\psi(X)
\equiv (- \partial_\mu \partial^\mu - \widehat m^2_z) \psi(X)\ ;\no
&&\widehat m^2_z = - \partial_z^2 + \varphi^2 (z) - \gamma_5 \varphi'(z) =
\widehat m^2_{+} P_L + \widehat m^2_{-} P_R\ , \la{f-s}
\ea
where  $P_{L,R} = \frac12 (1 \pm \gamma_5)$ are projectors on
the left- and right-handed states.
Thus the mass operator consists of two chiral {\sl superpartners} --
in the sense of
supersymmetric quantum mechanics 
%\ci{susy1,susy2,susy3}
\ci{susy1}-\ci{susy3} 
\ba
\widehat m_\pm^2 &=&  - \partial_z^2 + \varphi^2 (z) \mp \varphi'(z)
=  [\,-\partial_z \pm \varphi(z)\,][\,\partial_z \pm \varphi(z)\,]\ ;
\la{fact}\\
\widehat m_{+}^2\,q^+ &=& q^+\,\widehat m_{-}^2,\quad
\widehat m_{-}^2\,q^- = q^-\,\widehat m_{+}^2\ ,\quad q^\pm \equiv
\mp \partial_z + \varphi(z)\ . \la{susy}
\ea
The factorization \gl{fact} guarantees the positivity
of the mass operator -- {\it i.e.} the absence of tachyons --
and the supersymmetry realized
by the intertwining relations \gl{susy} entails the equivalence of the 
spectra between different chiralities
for non-vanishing masses. As a consequence, the related
left- and right-handed spinors can be assembled into the bi-spinor describing
a massive Dirac particle which, however, is not necessarily  localized at any
point of the extra-dimension if the 
field configuration $\varphi(z)$ is asymptotically constant. Indeed the massive
states will typically belong to the continuous spectrum -- or to a Kaluza-Klein
tower for the compact fifth dimension -- and spread out
the whole extra-dimension.
Meanwhile,
the spectral equivalence may be broken just by one single state, {\it i.e.}
a proper normalizable zero mode of
one of the mass operators $\widehat m_\pm^2$. Let us assume to get it in the
spectrum of $\widehat m_{+}^2$: then from Eq.s\ \gl{fact} and \gl{susy}
it follows that
\be
q^-\psi^{+}_0(x,z) = 0\ , \quad \psi^{+}_0(x,z) =
\psi_L(x) \, \exp\left\{-\int^z_{z_0} dw\varphi(w)\right\}\ ,
\ee
where $\psi_L(x) = P_L \psi (x)$ is a free
Weyl spinor in the four-dimensional Minkowski space.
Evidently, if a scalar field configuration has
the following asymptotic behavior: namely,
$$
\varphi(z)\stackrel{z \rightarrow \pm\infty}{\sim}
\pm C_\pm |z|^{\nu_\pm}\ ,\quad \mbox{\rm Re} \nu_\pm > -1\ ,
\quad  C_\pm > 0\ ,
$$
then the wave function $\psi^{+}_0(x,z)$ is normalizable 
on the $z$ axis and the corresponding
left-handed fermion is a massless Weyl particle localized in the vicinity
of a four-dimensional domain wall. It is easy to check that in this case
the superpartner mass operator  
$\widehat m_{-}^2$ does not possess a massless
normalizable solution and if
$\varphi(z)$ is asymptotically constant, with $ C_\pm > 0$ and $\nu_\pm = 0$,
there is always a gap for the massive Dirac states. Further on
we restrict ourselves to this scenario.

The important example of such a topological configuration can be derived 
for the system having the free mass spectrum -- continuum or Kaluza-Klein
tower -- for one of the chiralities, say, for right-handed fermions.
This is realized by
a kink-like scalar field background
\be
\varphi^{+} =
M\, \mbox{\rm tanh}(Mz)\ . \la{soli}
\ee 
The two mass operators have the following potentials
\be
\widehat m^2_{+} =- \partial_z^2 + M^2 
\left[\,1-2{\rm sech}^2(Mz)\,\right];\quad
\widehat m^2_{-} =- \partial_z^2 + M^2, \la{chpot}
\ee
and the left-handed normalized zero-mode is properly localized around $z=0$,
in such a way that we can set
\be
\psi^{+}_0(x,z) = 
\psi_L(x)\,\psi_0 (z)\ ,\qquad \psi_0 (z) \equiv
\sqrt{M/2}\ {\rm sech}(Mz)\ . \la{locmod}
\ee
As a consequence,
the threshold for the continuum is at $ M^2$
and the heavy Dirac particles may have
any masses $m > M$, the corresponding
wave-functions being completely de-localized in the extra-dimension.

On the one hand, 
if we investigate the scattering and decay of particles with energies
considerably smaller than $M$, we never reach the threshold of creation of
heavy Dirac particles with   $m > M$ and all physics interplays on the
four-dimensional domain wall with thickness $\sim 1/M$. 
On the other hand, at
extremely high energies, certain heavy fermions will appear in and
disappear from our world. 

It turns out that the real fermions -- quarks and leptons living on 
the domain wall
by assumption -- are mainly massive. Therefore, for each light fermion
on a brane one needs
at least two five-dimensional proto-fermions $\psi_1(X), \psi_2(X)$
in order to generate left- and right-handed
parts of a four-dimensional Dirac bi-spinor as zero modes.
Those fermions have clearly to couple with opposite charges
to the  scalar field $\Phi(X)$,
in order to produce the required zero modes with different chiralities when
$\langle\Phi(X)\rangle_0 = \varphi^+(z)$: namely,
\be
[\,i\not\!\partial - \tau_3\Phi(X)\,]\Psi(X) = 0\ ,\quad
\not\!\partial \equiv \widehat\gamma_\alpha \partial^\alpha\ ,\quad
\Psi(X) =\left\lgroup\begin{array}{c}\psi_1(X)\\ 
\psi_2(X)\end{array}\right\rgroup\ ,
\la{2fer}
\ee
where $\widehat\gamma_\alpha \equiv \gamma_\alpha\otimes {\bf 1}_2$ are
Dirac matrices and 
$\tau_a \equiv {\bf 1}_4 \otimes \sigma_a,\ a=1,2,3 $
are the generalizations of
the Pauli matrices $\sigma_a$ acting on the bi-spinor components $\psi_i(X)$.

In this way one obtains a massless Dirac particle on the brane 
and the next task 
is to supply it with a light mass. As the mass operator
mixes left- and right-handed
components of the four-dimensional fermion it is embedded in the Dirac 
operator \gl{2fer} with the mixing matrix $\tau_1 m_f$ of the fields
$\psi_1(X)$ and $\psi_2(X)$. 
At last, following the general Standard Model mechanism
of fermion mass generation by means of the Higgs scalars, 
one can introduce the
second scalar field $H(x)$ to make this job, replacing the bare mass
$\tau_1 m_f \longrightarrow \tau_1 H(x)$. Both scalar fields might
be dynamical indeed and their self-interaction should justify the spontaneous
symmetry breaking by certain classical configurations 
allocating light massive fermions on the domain wall.
From the previous discussion it follows that the minimal
set of five-dimensional fermions has to include 
two Dirac fermions coupled to
scalar backgrounds of opposite signs.
In addition to the trapping scalar field,
another one is in order to supply light domain wall fermions
with a mass. Thus we introduce two types of four-fermion self-interactions
to reveal two composite scalar fields with  a proper coupling to fermions.
As we shall see, these two scalar fields will
acquire mass spectra similar to fermions with light counterparts located
on the domain wall. The dynamical mechanism of creation of domain wall
particles turns out to be quite economical and few predictions on masses
and decay constants of fermion and boson particles will be derived.

%%%%%%%%%%%%%%%%%%%%%%%%%%%%%%%%%%%%%%%%%%%%%%%%%%%%%%%%%%%%%%%%%%%%%%%%%%%%%%%%

\section{Fermion model with self-interaction in 5D}
%five-dimensions}

Let us consider the model with the
following Lagrange density
\be
{\cal L}^{(5)} (\overline{\Psi},\Psi) = \overline{\Psi}\ i \!\not\!\partial 
\Psi +
\frac{g_1}{4N \Lambda^3}\left(\overline{\Psi}\tau_3 \Psi\right)^2  
+ \frac{g_2}{4N \Lambda^3}\left(\overline{\Psi} \tau_1\Psi\right)^2,  
\la{mod}
\ee 
where
$\Psi(X)$ is an eight-component five-dimensional
fermion field, see Eq.\,\gl{2fer} --
either a bi-spinor in a four-dimensional theory or a spinor in a
six-dimensional theory --
which may also realize a flavor and color multiplet with the total number
$N=N_f\,N_c$ of states. The ultraviolet
cut-off scale $\Lambda$ bounds fermion momenta, as the four-fermion interaction is
supposed here to be an effective one, 
whereas $g_1$ and $g_2$ are suitable dimensionless
and eventually scale dependent effective couplings.

This Lagrange density can be more transparently treated with the help of a pair of
auxiliary scalar fields $\Phi(X)$ and $H(X)$, which eventually will allow to
trap a light fermion on the domain wall and to supply it with a mass
\ba
&&{\cal L}^{(5)} (\overline{\Psi},\Psi,\Phi, H) =\no
&&\overline{\Psi} ( i\!\not\!\partial
- \tau_3 \Phi -\tau_1 H) \Psi - \frac{N \Lambda^3}{g_1}\,\Phi^2
- \frac{N \Lambda^3}{g_2}\,H^2\ . \la{aux}
\ea
In this model the invariance (when it holds) under discrete $\tau$-symmetry
transformations
\ba
&&\Psi \longrightarrow \tau_1 \Psi\ ;\quad
\Phi \longrightarrow -\,\Phi\ ;\\
&&\Psi \longrightarrow \tau_2 \Psi\ ;\quad
\Phi, H \longrightarrow -\,\Phi, -\,H\ ;\\
&&\Psi \longrightarrow \tau_3 \Psi\ ;\quad
H \longrightarrow -\,H\ ,
\la{tausim}
\ea
does not allow the fermions to acquire
a mass and prevents a breaking of translational invariance
in the perturbation theory.
This $\tau$-symmetry can be associated to the chiral symmetry in four
dimensions \footnote{ It can
be also related to the chiral symmetry in a six-dimensional space-time
where from our five-dimensional model can be derived by 
dimensional reduction.}.

However for sufficiently strong couplings,
this system undergoes
the phase transition to the state  in which the condensation of
fermion-anti-fermion pairs does spontaneously  break --
partially or completely -- the $\tau$-symmetry.

The physical meaning of the scale $\Lambda$
is that of a compositeness scale for heavy scalar bosons
emerging after the breakdown of the $\tau$-symmetry.
In order to develop the infrared phenomenon of $\tau$-symmetry breaking,
the effective Lagrange density containing the essential low-energy
degrees of freedom has to be derived.

To this concern we proceed -- only in this Section --
to the transition to the Euclidean space,
where the invariant four-momentum cut-off can be unambiguously implemented.
Within this framework, the notion
of low-energy is referred to momenta $|p| < \Lambda_0$ as compared to
the cut-off $\Lambda \gg \Lambda_0$. However, after the elaboration 
of the domain wall vacuum, we will search for the fermion states 
with masses $m_f$ much lighter than the dynamic scale $\Lambda_0$,
{\it i.e.} for the ultralow-energy physics.
Thus, eventually, there are three scales in the present model
in order to implement the domain wall particle trapping.

Let us decompose the momentum space fermion fields into their high-energy part
$\Psi_h(p)\equiv\Psi(p)\vartheta(|p|-\Lambda_0)\vartheta(\Lambda-|p|)$,
their low-energy part $\Psi_l(p)\equiv\Psi(p)\vartheta(\Lambda_0-|p|)$
and integrate out the high-energy part of the fermion spectrum,
$\vartheta(t)$ being the usual Heaviside step distribution.
More rigorously, the above decomposition of the fermion spectrum
should be done covariantly, {\it i.e.}
in terms of the full Euclidean Dirac operator,
\be
D \equiv i (\not\!\partial + \tau_3 \Phi + \tau_1 H)\ .\la{eudir}
\ee 
Nevertheless, as we want to concentrate ourselves on the triggering of
$\tau$-symmetry breaking by fermion condensation, we can safely assume to
neglect further on the high-energy components of the auxiliary boson fields,
what is equivalent to the use of the mean-field or large $N$ approximations.
Then the low-energy Lagrange density, which solely accounts for
low-energy fermion and boson fields, can be written as the sum of
the expression in Eq.\
\gl{aux} and the one-loop effective Lagrange density: namely,
\be
{\cal L}^{(5)}_{\rm low} (\overline{\Psi}_l,\Psi_l,\Phi, H) =
{\cal L}^{(5)}_E (\overline{\Psi}_l,\Psi_l, \Phi, H) +
\Delta {\cal L}^{(5)} (\Phi, H), \la{lowl}
\ee
where the tree-level low-energy Euclidean Lagrange density reads
\be
{\cal L}^{(5)}_E (\overline{\Psi}_l,\Psi_l,\Phi, H) =
i\overline{\Psi}_l (\!\not\!\partial
+ \tau_3 \Phi +\tau_1 H) \Psi_l + \frac{N \Lambda^3}{g_1}\,\Phi^2
+ \frac{N \Lambda^3}{g_2}\,H^2.
\ee
%\ea
The one-loop contribution of high-energy fermions is given by
\ba
&&\Delta {\cal L}^{(5)} (\Phi, H)=
C-(N/2)\,\mbox{tr}\langle X|{\rm A}|X\rangle\la{leff}\ , \\
&&{\rm A} \equiv
\vartheta(\Lambda^2 - D^\dagger D)
\ln\frac{D^\dagger D}{\Lambda^2}
- \vartheta(\Lambda_0^2 - D^\dagger D)
\ln\frac{D^\dagger D}{\Lambda_0^2}\ ,
\nonumber
\ea
where the normalization constant $C$ is such that
$\Delta {\cal L}^{(5)} (0, 0) = 0$ and the
${\rm tr}$ operation stands for the trace
over spinor and internal degrees of freedom.
In Eq.\,\gl{leff}
the choice of normalizations in the logarithms ensures the continuity 
of spectral
density  at the positions of cut-offs $\Lambda$ and $\Lambda_0$.
Thereby the spectral flow through the
spectral boundary is suppressed.
In the latter operator ${\rm A}$ we have incorporated the cut-offs
which select out the high-energy region defined above \ci{AnBo}.\\
\noindent
From the conjugation
property $D^\dagger = \tau_2 D \tau_2$, it follows that
the Euclidean Dirac operator
is a normal operator, which has to be implemented in order to get
a real effective action
and to define the spectral cut-offs with the help of the positive operator
\ba
D^\dagger D &=& -\partial_\mu\partial_\mu - \partial_z^2 + {\cal M}^2(X)
=-\partial_\alpha\partial_\alpha+ {\cal M}^2(X)\ ,\no
{\cal M}^2(X) &\equiv&
\Phi^2(X)  + H^2(X)   - \tau_3 \not\!\partial \Phi(X)
 - \tau_1 \not\!\partial H(X)\ . \la{quad}
\ea
One can
see that, in fact, the scale anomaly only contributes 
into  $\Delta {\cal L}^{(5)}$, {\it i.e.} that part which depends on the scales. 
Thus, equivalently,
\be
 \Delta {\cal L}^{(5)} (\Phi, H) = C + N \int^{\Lambda}_{\Lambda_0}
\frac{dQ}{Q}\
\mbox{tr}\langle X|
\vartheta(Q^2 - D^\dagger D)|X\rangle\ .
\la{scan}
\ee
As we assume that the scalar fields carry momenta much smaller than both scales
$\Lambda \gg \Lambda_0$, then the diagonal matrix element
in the RHS of Eq.\,\gl{scan} can be
calculated either with the help of the derivative expansion of the master
representation \ci{AnBo}
\ba
&&\langle X|\vartheta(Q^2 - D^\dagger D)|X\rangle =Q^n
\int_{-\infty}^{+\infty}\frac{d t}{2\pi i}\ 
\frac{\exp\{i t\}}{t - i\varepsilon}
\no
&&\times \int \frac{d^n k}{(2\pi)^n} \exp\left\{- \frac{i t}{Q^2}
\left[\,k^2 + 2iQ
k_\alpha \partial_\alpha - \partial_\alpha\partial_\alpha +
{\cal M}^2(X)\,\right] \right\}, \la{intrep}
\ea
where $n$ is a number of dimensions of the Euclidean space,
or by means of the heat kernel asymptotic expansion.
For $n\le 5 $, only three heat kernel coefficients at most
are proportional to non-negative powers of the large parameter $Q$ 
(see Appendix A)
\ba
&&\langle X|\vartheta(Q^2 - D^\dagger D)|X\rangle = C_0 Q^n + 
C_1 Q^{n-2} \,{\cal M}^2(X) \no
&& + Q^{n-4}  
\left\{C_2\,[{\cal M}^2(X)]^2+C_3\,\partial_\alpha\partial_\alpha
[{\cal M}^2(X)]
\right\} + {\cal O}\left(Q^{n-6}\right), \la{elem}
\ea
where, for $n < 6$ and large scales $\Lambda_0 < Q < \Lambda$, 
the neglected terms
rapidly vanish. For the given operator $\vartheta(Q^2 - D^\dagger D)$,
the coefficients $C_i$ take the following values
\ba
&&C_0= \frac{1}{n\ 2^{n-1}\,\pi^{n/2}\,\Gamma(n/2)}\ ;\no
&&C_1= - \frac{1}{ 2^{n}\,\pi^{n/2}\,\Gamma(n/2)}
= - \frac{n}{2}\,C_0 < 0\ ;\no
&&C_2 = \frac{n-2}{ 2^{n+2}\,\pi^{n/2}\,\Gamma(n/2)}
= \frac{n(n-2)}{8}\,C_0 > 0\ ;\no
&&C_3 = -\frac{n-2}{3\cdot 2^{n+2}\,\pi^{n/2}\,\Gamma(n/2)}
= -\frac{n(n-2)}{24}\,C_0 < 0\ ,
\ea 
where $\Gamma(y)$ stands for the Euler's gamma function. For different
possible definitions of the effective Lagrange density, involving 
operators and regularization functions other than those ones
of Eq.\,\gl{scan},
one might obtain in general different sets of $C_i$,  albeit
their signs are definitely firm.
As we shall see further on, 
the negative sign of $C_1$ catalyzes the chiral symmetry
breaking at sufficiently strong coupling constants.
Taking the little trace and integrating the RHS of Eq.\,\gl{elem},
from Eq.\,\gl{scan} one finds, up to a total $n$-divergence and
setting $C=C_0(\Lambda_0^n-\Lambda^n)N/n$,
\ba
\Delta {\cal L}^{(5)} (\Phi, H)\stackrel{\Lambda\rightarrow\infty}{\sim}
 &-& A_1 \left(\Lambda^{n-2} - \Lambda_0^{n-2}\right)
\left(\Phi^2 + H^2\right)\no
&+&  A_2 \left(\Lambda^{n-4}- \Lambda_0^{n-4}\right)
\left[\,(\partial_\alpha \Phi)^2 +
(\partial_\alpha H)^2\,\right]\no
&+&  A_2 \left(\Lambda^{n-4}- \Lambda_0^{n-4}\right)
\left[\,\left(\Phi^2 + H^2\right)^2\,\right]\ ,
\la{acti}
\ea
where
\ba
A_1 &=& \frac{N}{(n-2) 2^{n-3}\,\pi^{n/2}\,\Gamma(n/2)}
\stackrel{n\uparrow 5}{\longrightarrow} \frac{N}{9\pi^3}\ ;\\
A_2 &=& \frac{N (n-2)}{(n-4) 2^{n-1}\,\pi^{n/2}\,\Gamma(n/2)}
\stackrel{n\uparrow 5}{\longrightarrow} \frac{N}{4\pi^3}\ .
\ea
As $\Lambda_0 \ll \Lambda$ one can neglect the $\Lambda_0$-dependence
in Eq.\,\gl{acti} for $n > 4$,
whereas near four dimensions the pole in $A_2$ together with the cut-off 
dependent factor generate the coefficient $\sim 
\ln\left(\Lambda/\Lambda_0\right)$.
For $n=5$ we eventually find
\ba
&&\Delta {\cal L}^{(5)} (\Phi, H)\stackrel{\Lambda\rightarrow\infty}{\sim}\no
&&\frac{N\Lambda}{4\pi^3}\left[\,(\partial_\alpha \Phi)^2 +
(\partial_\alpha H)^2\,\right]-\frac{N\Lambda^3}{9\pi^3}\left(\Phi^2 + 
H^2\right)
+\frac{N\Lambda}{4\pi^3}\left(\Phi^2 + H^2\right)^2\ .\la{a12}
\ea
Although the actual values of the coefficients $A_1$ and $A_2$ might
be regulator-de\-pen\-dent,
as already noticed,
the coefficients of the kinetic
and quartic terms of the effective Lagrange density are definitely equal,
no matter how the latter is 
obtained from the basic Dirac operator of Eq.\,\gl{eudir}.
This very fact is at the origin of the famous Nambu relation between
the dynamical mass of a fermion and the mass of a scalar bound state in
the regime of $\tau$-symmetry breaking, as we will see in the next Section.
%%%%%%%%%%%%%%%%%%%%%%%%%%%%%%%%%%%%%%%%%%%%%%%%%%%%%%%%%%%%%%%%%%%%%%%%%%%%%%%%%%%
\section{$\tau$-symmetry breaking}

The interplay between different operators in the low-energy
Lagrange density \gl{lowl} may lead to two different dynamical
regimes, depending on whether the $\tau$-symmetry
is broken or not.
Indeed, going back to the Minkowski five-dimensional space-time,
the low-energy Lagrange density can be suitably cast in the form
\ba
{\cal L}_{\rm low}^{(5)} (\overline{\Psi}_l,\Psi_l,\Phi, H) &=&
\overline{\Psi}_l(X) [\,i\!\not\!\partial
- \tau_3 \Phi(X) -\tau_1 H(X)\,] \Psi_l(X)\no
&+& \frac{N\Lambda}{4\pi^3}\,\partial_\alpha \Phi(X)\partial^\alpha \Phi(X)+
\frac{N\Lambda}{4\pi^3}\,\partial_\alpha H(X)\partial^\alpha H(X)\no
&+& \left(\frac{N}{9\pi^3} - \frac{N}{g_1}\right)
\Lambda^3 \Phi^2(X)
+\left(\frac{N}{9\pi^3} - \frac{N}{g_2}\right)\Lambda^3 H^2(X)\no
&-& \frac{N\Lambda}{4\pi^3}\,[\,\Phi^2(X) + H^2(X)\,]^2. \la{lowmin}
\ea
We notice that
the quartic operator in the potential is symmetric and positive, whilst
the quadratic operators have different signs for weak and strong
couplings, the critical values for both couplings being the same
within the finite-mode cut-off regularization of Eq.\,\gl{a12}: namely,
\be
g_i^{\rm cr} = 9\pi^3\ ,\qquad i=1,2\ .
\ee
Let us introduce two mass scales $\Delta_i$ in order to parameterize the
deviations from the critical point
\be
\Delta_i(g_i)
%\frac{\Lambda^2}{2 A_2}\left(A_1 - \frac{N}{g_i}\right)
=\frac{2\Lambda^2}{9g_i}\left(g_i-g_i^{\rm cr}\right)\ .
\ee
Taking into account that $g_i\ge 0$ we have
\be
\Delta_i(g_i) \leq (2/9)\Lambda^2
\la{delta}
\ee
and the effective Lagrange density for the scalar fields 
takes the simplified form
\be
{\cal L}^{(5)}_{\rm scalar} (\Phi, H)  = \frac{N\Lambda}{4\pi^3}
\left[(\partial_\alpha \Phi)^2 +
(\partial_\alpha H)^2 + 2\Delta_1 \Phi^2 + 2\Delta_2 H^2 -
(\Phi^2 + H^2)^2 \right]\ . \la{slag}
\ee

So far two constants $g_i$, and thereby $\Delta_i$,
play an equivalent role and the related vertices are invariant under the
replacements $\Psi_l\longmapsto\tau_2\Psi_l$ and subsequent
reflection of the scalar fields -- see Eq.\,\gl{tausim}.
Therefore, without loss of generality, one can always choose
\be
\Delta_1(g_1) > \Delta_2(g_2)\ .
\ee
Whenever both couplings $g_i$ are within the range $0<g_i<g_i^{\rm cr}= 9\pi^3$,
then we have $\Delta_i(g_i)< 0$ and consequently
the potential has a unique symmetric minimum.
If instead one at least of the couplings  $g_i$ does
exceed its critical value, then the symmetric
extremum at $\Phi = H = 0$ is no longer a minimum, though either a saddle
point for $\Delta_2(g_2) < 0 < \Delta_1(g_1)$ or even a maximum for
$0<\Delta_2(g_2) < \Delta_1(g_1)$.
If $\Delta_1(g_1) > 0$, then the true minima appear at a non-vanishing
vacuum expectation value of the scalar field $\Phi(X)$: namely,
\be
(I)\qquad
\quad \Phi_I \equiv \langle\Phi(X)\rangle_0= \pm \sqrt{\Delta_1(g_1)}\ ,
\quad H_I\equiv  \langle H(X)\rangle_0 = 0\ .\la{consol}
\ee
This follows from the stationary point conditions for constant fields
\ba
&&\left[\Delta_1(g_1) - \Phi^2 - H^2\right]\Phi=0\ ,\no
&&\left[\Delta_2(g_2) - H^2 - \Phi^2\right]H=0
\la{exeq}
\ea
and from the positive definiteness of the second variation
of the boson effective action for constant boson fields.
As a matter of fact, if we set
\ba
S&\equiv&(S_1,S_2)=(\Phi,H)\ ,\qquad
S_I\equiv\left(\pm \sqrt{\Delta_1},0\right)\ ,\no
{\cal V}[S]&\equiv& \frac{N\Lambda}{4\pi^3}\int d^5 X\
\left\{ - 2\Delta_i\,S_i^2(X) +
[S_i(X)S_i(X)]^2 \right\}\ ,\la{poten}
\ea
we readily find
\ba
{\cal V}[S]-{\cal V}\left[S_I\right] &=&
\frac{1}{2}\int d^5X\int d^5Y\ s_i(X)s_j(Y)\left.
\frac{\delta^{2}{\cal V}[S]}{\delta S_i(X)\delta S_j(Y)}\right|_{S=S_I}
+\ldots\ ,\no
&=&\frac{N\Lambda}{4\pi^3}\int d^5X\
s_i(X)s_j(X)\left\lgroup{\bf M}_I\right\rgroup_{ij}
+\ldots\ ,\label{der2}
\ea
where
\be
s_i(X)\equiv S_i(X)-S_I\ ,\qquad i=1,2,\label{mistero}
\ee
whereas the Hessian mass matrix ${\bf M}_I$:
\be
{\bf M}_I =
\left\lgroup\begin{array}{cc}
4\Delta_1 & 0 \\
0 & 2\Delta_1 - 2\Delta_2
\end{array}\right\rgroup \equiv \left\lgroup\begin{array}{cc}
 M_1^2 & 0 \\
0 &   M_2^2
\end{array}\right\rgroup\ ,\la{hess}
\ee
is manifestly positive definite and determines the mass spectrum
of the five-di\-men\-si\-onal scalar excitations.

A further constant solution of Eq.\,\gl{exeq} does exist for $\Delta_2> 0$,
{\it i.e.} $\langle\Phi\rangle_0=  0$ and
$\langle H\rangle_0 = \pm \sqrt{\Delta_2} $.
However, it corresponds to a saddle point of the potential,
as it can be seen from Eq.\,\gl{der2} for
$\Delta_1 > \Delta_2$. Likewise, if $\Delta_1 > \Delta_2 >0 $, then
 the matrix   ${\bf M}$ is
 negative definite at the symmetric point
$\langle\Phi\rangle_0 = \langle H\rangle_0 = 0$
which corresponds thereby to a maximum.
The degenerate situation -- {\it i.e.} the valley -- actually
occurs for $\Delta_1 = \Delta_2 >0 $, when the rotational
$\tau_2$-symmetry is achieved by the Lagrange density
but is spontaneously broken.
The massless scalar state in Eq.\,\gl{hess} arises
in full accordance with the Goldstone's theorem.

The corresponding dynamical effect for the fermion model of Eq.\,\gl{mod}
consists in the formation of a
fermion condensate and the generation of a dynamical fermion mass
$M$ -- see Eq.s\ \gl{aux} and \gl{lowl} -- that breaks
the $\tau_2$-symmetry. Its ratio to the heaviest scalar mass just obeys
the Nambu relation
\be
 M_1\equiv 2M > M_2\ , \qquad
M \equiv \langle\Phi\rangle_0= \sqrt{\Delta_1}\ ,
\ee
the second, lighter composite scalar being a pseudo-Goldstone state.
We notice that
the above relationship holds true independently of the specific
values of the coefficients $A_i$
in Eq.\,\gl{acti} and, consequently, it takes place in four and five dimensions.
However, if we properly re-scale the scalar fields according to
\be
\Phi(X)\sqrt\Lambda\equiv \pm M\sqrt\Lambda+\nu\,u(X)\ ,\qquad
H(X)\sqrt\Lambda \equiv\nu\,\upsilon(X)\ ,\qquad
\nu\equiv\sqrt{2\pi^3/N}\ ,
\la{rescale}
\ee
in such a way that $\dim[u]=\dim[\upsilon]=3/2$,
then the low-energy Lagrange density \gl{lowmin}
can be suitably recast in the form
\ba
{\cal L}_{\rm low}^{(5)} (\overline{\Psi}_l,\Psi_l,u,\upsilon) 
&=&M^4\Lambda/2\nu^2+
i\overline{\Psi}_l(X)\!\not\!\partial\Psi_l(X) - 
M\overline{\Psi}_l(X)\tau_3\Psi_l(X)
\no
&+&
(1/2)\partial_\alpha u(X)\partial^\alpha u(X) - 2M^2 u^2(X)\no
&+&
(1/2)\partial_\alpha \upsilon(X)\partial^\alpha \upsilon(X)
- 2(M^2-\Delta_2)\upsilon^2(X)\no
&-&
\nu\Lambda^{-1/2}\left[\,\overline{\Psi}_l(X)\tau_3\Psi_l(X) u(X) -
\overline{\Psi}_l(X)\tau_1\Psi_l(X)\upsilon(X)\,\right]\no
&\mp&
2\nu M\Lambda^{-1/2}\ [\,u^2(X)+\upsilon^2(X)\,]u(X)\no
&-&(\nu^2/2)\Lambda^{-1}\ [\,u^2(X)+\upsilon^2(X)\,]^2\ .
\ea
As a consequence, one can see that all the low-energy effective couplings
for fermion and boson fields do rapidly vanish in the large cut-off limit.
In particular, if the cut-off $\Lambda$ is much
larger than the energy range of our physics, then we are dealing 
with a theory of
practically free, non-interacting particles.
Moreover, this pattern of $\tau$-symmetry breaking does not provide
the desired trapping on a domain wall:
the heavy fermions and bosons
live essentially in the whole five-dimensional space.
From now on we shall proceed to consider another type of vacuum solutions,
which break the five-dimensional translational invariance and
give rise to the formation of domain walls.

%%%%%%%%%%%%%%%%%%%%%%%%%%%%%%%%%%%%%%%%%%%%%%%%%%%%%%%%%%%%%%%%%%%%%%%%%%%%%%%%%%%%

\section{Domain walls: massless phase}
The existence of two minima in the potential of Eq.\,\gl{poten} gives rise
to another
set of vacuum solutions \ci{Rajar} - \ci{Baz}, 
which connect 
smoothly the minima owing to the kink-like
shape of Eq.\,\gl{soli} with $M = \sqrt{\Delta_1}$. 
On variational and geometrical
grounds one could expect that certain minimal solutions are collinear,
just breaking the translational invariance  in one direction.
We specify this direction along the fifth coordinate $z$. Then one can
discover two types of competitive solutions \ci{Mac}-\ci{Baz},
\ba
(J)\qquad\quad &&\langle\Phi(X)\rangle_0 \equiv
\Phi_J (z) =\pm M
\mbox{\rm tanh}(M z)\ , \no
&&  \langle H(X)\rangle_0\equiv H_J (z)  = 0\ ;\la{Jvacuum}\\
(K)\qquad\quad &&\langle\Phi(X)\rangle_0\equiv \Phi_K (z) 
= \pm M
\mbox{\rm tanh}(\beta z)\ ,\no &&
\langle H(X)\rangle_0 \equiv H_K (z)  =\pm\mu\,\mbox{\rm sech}(\beta z)\ .
\la{Kvacuum}
\ea
Further on we select out only positive signs in vacuum configurations 
to analyze the scalar
fluctuations around them, having in mind that our analysis is absolutely 
identical around other configurations.
When we insert the second solution $(K)$ into the stationary point
conditions
\ba
&&2\left[\,M^2 - \Phi^2 - H^2\,\right]\Phi=
\partial^\alpha\partial_\alpha\Phi\ ,\no
&&2 \left[\,\Delta_2 - H^2 - \Phi^2\,\right]H=
\partial^\alpha\partial_\alpha H\ ,
\la{steq}
\ea
we find
\be
\mu =
\sqrt{2 \Delta_2 - M^2}\ ,\qquad
\beta = \sqrt{M^2 - \mu^2}\ . \la{mu}
\ee
The solution $(K)$ exists only for $\Delta_2< M^2 < 2\Delta_2$
and it coincides with the extremum $(J)$ in the limit
$\Delta_2\to M^2/2,\ \mu\to 0,\ \beta\to M$.
The question arises
about which one of the two solutions is a true minimum and
whether they could coexist if $\mu > 0 $.
The answer can be obtained from the analysis of the second variation
of the bosonic low-energy effective action.
The corresponding relevant second order differential operator
can be suitably written in terms of the notations introduced in
Eq.\,\gl{poten}: namely,
\be
\left\lgroup{\bf D}^{2}_X\right\rgroup_{ij}\equiv -\ \delta_{ij}\Box_x
-\left\lgroup{\bf M}[S(X)]\right\rgroup_{ij}\ ,\qquad
\Box_x\equiv\partial^\mu\partial_\mu\ ,
\ee
\be
\left\lgroup{\bf M}[S(X)]\right\rgroup_{ij}\equiv\delta_{ij}
\left[- \partial_z^2-2\Delta_i + 2 S_k(X) S_k(X)\right]+4S_i(X)S_j(X)\ .
\label{massa}
\ee
The stationary point solutions $(J)$ and $(K)$ are true minima iff
the matrix-valued mass operator ${\bf M}[S(X)]$
becomes positive semi-definite at the extrema
\be
S_{J}=\left( M\mbox{\rm tanh}M z,0\right)\ ,\qquad
S_{K}=\left( M\mbox{\rm tanh}\beta z,
 \mu\,\mbox{\rm sech}\beta z\right)\ .
\ee
Now, at the stationary point $(J)$ the matrix-valued mass operator
\be
{\bf M}_{J}(z)\equiv{\bf M}\left[S_{J}(z)\right]
\ee
turns out to be diagonal with entries
\ba
{\rm M}_{J}^{11}&\equiv&
- \partial_z^2 + 4M^2 - 6M^2\,{\mbox{\rm sech}^2(M z)}\ ,\label{massII1}\\
{\rm M}_{J}^{22}&\equiv&
- \partial_z^2 +2M^2 - 2 \Delta_2
-2M^2\,{\mbox{\rm sech}^2(M z)}\ ,\label{massII2} \\
{\rm M}_{J}^{12}&=&{\rm M}_{J}^{21} = 0.
\ea
Both components do represent one-dimensional Schr\"odinger-like operators,
the eigenvalue problem of which can be exactly solved analytically.
The Schr\"odinger-like operators \gl{massII1} and \gl{massII2}
can be presented in the factorized form similar to that one of
Eq.\,\gl{fact}: namely,
\ba
&&{\rm M}_{J}^{11}  =
\left[- \partial_z + 2M\mbox{\rm tanh}(M z)\right]
\left[\partial_z + 2M\mbox{\rm tanh}(M z)\right]\ ;\label{factor1}\\
&&{\rm M}_{J}^{22} = M^2- 2 \Delta_2
+ \left[- \partial_z + M\mbox{\rm tanh}(M z)\right]
\left[\partial_z + M\mbox{\rm tanh}(M z)\right]\ \label{factor2}
\ea
Therefrom it is straightforward to check that
the ground states of the operator ${\bf M}_{J}(z)$ are described by the
real, node-less in $z$ and normalized wave functions
\ba
&&
{\rm M}_{J}^{11}\,\phi_J (z) =0\ ,\quad
{\rm M}_{J}^{22}\, h_J (z)= m^2_h\,h_J(z)\ ,
\quad m^2_h \equiv M^2 - 2\Delta_2\  ;
\label{ground0}\\
&&
\phi_J(z) \equiv
{\mbox{\rm sech}^2(M z)}\sqrt{3M/4}\ ,\quad
h_J(z)\equiv {\mbox{\rm sech}(M z)}\sqrt{M/2}\ ;
\label{ground1}
\ea
in such a way that we can suitably parameterize the shifts of
the scalar field with respect to the background vacuum solution
$(J)$ -- see Eq.s\,\gl{mistero} and  \gl{Jvacuum} -- by the following
two eigenstates of the mass matrix ${\bf M}_{J}(z)$: namely,
\be
\Omega^\phi_{J}(X) = \phi(x)\left\lgroup\begin{array}{c}
\phi_J(z)\\
0\end{array}\right\rgroup\frac{\nu}{\sqrt\Lambda}\ , \qquad
\Omega^h_{J}(X) = h(x)\left\lgroup\begin{array}{c}
0\\
h_J(z)\end{array}\right\rgroup\frac{\nu}{\sqrt\Lambda}\ ,\label{ground2}
\ee
where $\phi(x),h(x)$ do eventually represent the ultralow-energy 
scalar fields
on the Minkowski space-time, as we shall better see  below on.
As a consequence, the spectrum of the second variation is positive
if $M^2 > 2\Delta_2$ and, in this case, the solution $(K)$
does not exist whilst the scalar lightest states are localized
on the domain wall. More precisely -- see Appendix B --
the first boson $\Phi$ has two states on the brane: 
a massless state and
a heavy massive state of mass $\sqrt{3M}$. The existence
of  a massless scalar state around the kink
configuration $(J)$ is a consequence of the spontaneous breaking
of translational
invariance -- see next Section.  Other heavy states belong
to the continuous part of the spectrum with threshold
at $2 M$.
The second boson $H$ has only one state on the brane of mass
$\sqrt{M^2 - 2\Delta_2}$ and its continuous
spectrum starts at $\sqrt{2M^2- 2\Delta_2} \geq M$.

Since the vacuum expectation value
of the scalar field  $\langle\Phi(X)\rangle_0=M
\mbox{\rm tanh}(M z)$
has a kink shape, its coupling to fermions induces
the trapping of the lightest, massless fermion state on the domain wall: namely,
\be
\Psi_0 (X) = \left\lgroup\begin{array}{c}\psi_{1L}(x)\\
\psi_{2R}(x)\end{array}\right\rgroup\psi_0(z)\ ,\qquad
\psi_0(z)={\mbox{\rm sech}(M z)}\sqrt{M/2}\ ,
\la{psi0}
\ee
see Eq.s~\gl{locmod} and \gl{2fer}. The continuum of the heavy fermion states
begins at $M$ and involves pairs of heavy Dirac fermions.

In conclusion, at ultralow energies much smaller than $M$,
the physics in the neighborhood of the vacuum $(J)$ is essentially
four-dimensional in the fermion and boson sectors. 
It is described by the massless Dirac fermion
\be
\psi(x) = \left\lgroup\begin{array}{c}\psi_{L}(x)\\
\psi_{R}(x)\end{array}\right\rgroup\ ,
\ee
with $\psi_{L}(x)$ and $\psi_{R}(x)$ being
the two-component non-trivial parts  of $\psi_{1L}(x)$ and 
$\psi_{2R}(x)$ respectively, in such a way that we can set
\be
\Psi_l(X) = \left\lgroup\begin{array}{c}\psi_{L}(x)\\
\psi_{R}(x)\end{array}\right\rgroup\psi_0(z)\ .\la{psi5}
\ee
In result one has two four-dimensional scalar bosons,
a massless one and a massive one, provided
$M^2- 2\Delta_2 \ll M^2$ otherwise there is decoupling.

The matrix $\tau_3$ does not mix the two types of fermions
$\psi_{1L}$ and $\psi_{2R}$, but the related Yukawa vertex
in the Lagrange density \gl{aux}
mixes left-  and right-handed components
of each of them. As a consequence
the massless scalar field $\phi(x)$ does not couple directly
to a light fermion-anti-fermion pair. Its coupling to fermions
involves inevitably
heavy fermion degrees of freedom. Therefore, the ultralow-energy 
effective action  does not contain a Yukawa-type vertex 
for the field $\phi(x)$, which appears thereby to be {\it sterile}.

On the other hand, the
interaction between light fermions and the second scalar field $h(x)$ 
on the domain wall does achieve the conventional
Yukawa form.
Indeed, once they are projected on
the zero-mode space of $\psi(x)\,\psi_0(z)$, the matrices
$\tau_1$, $\tau_2$ and $\tau_3$ act as the Dirac 
matrices in the chiral representation: namely,
\be
\tau_1 \longrightarrow \gamma_0 =\left(\begin{array}{cc}
0 & \mbox{\bf 1}\\
\mbox{\bf 1} & 0\end{array}\right)\ ;\quad
\tau_3 \longrightarrow \gamma_5 = \left(\begin{array}{cc}
\mbox{\bf 1}& 0\\
0 & -\mbox{\bf 1}\end{array}\right)\ ;\quad
\tau_2 \longrightarrow i \gamma_0\gamma_5\ ;
\ee
where the $2\times2$ unit matrix ${\bf 1}$ acts on Weyl components
$\psi_{L}(x)$ and $\psi_{R}(x)$. Meanwhile the matrices in the kinetic part
of the Dirac operator are 
projected onto the zero-mode space as
\be
\widehat\gamma_0 \widehat\gamma_k \longrightarrow 
\left(\begin{array}{cc}
 \sigma_k& 0\\
0 & - \sigma_k\end{array}\right) \equiv  \gamma_0 \gamma_k\ ,
\ee
where the Pauli matrices $\sigma_k$ act on the two-component spinors 
$\psi_{L}(x)$ and $\psi_{R}(x)$.
As a consequence, the matrix $\tau_1 \longrightarrow \gamma_0$ just induces
the Yukawa mass-like vertex in the effective action 
on the domain wall.

As a final result, in the
vicinity of the vacuum solution $(J)$ of Eq.\,\gl{Jvacuum},
the ultralow-energy effective Lagrange density for the light
states on the four-dimensional Minkowski space-time comes out from
Eq.s\,\gl{lowmin},\gl{ground1},\gl{ground2},\gl{psi5} and reads
\footnote{To be precise, the ultralow-energy effective Lagrange density 
for the light states on the four-dimensional Minkowski space-time is
well defined up to subtraction of an infrared divergent constant.}
\ba
{\cal L}_{J}^{(4)}\left[\,\overline{\psi},\psi,\phi, h\,\right]
&=& \int^{+\infty}_{-\infty} dz\
{\cal L}_{\rm low}^{(5)}\left[\,\overline{\Psi}_l(X),\Psi_l(X),
\Omega_{J}^\phi(X),\Omega_{J}^h(X)\,\right]\\
\la{lagr4}
&=& i\overline{\psi}(x)\gamma^\mu \partial_\mu\psi(x)
+\frac12\partial_\mu\phi(x) \partial^\mu\phi(x)\no
&+&\frac12\partial_\mu h(x)\partial^\mu h(x)
-\frac12 m^2_h h^2(x)\no
&-&g_f\overline{\psi}(x)\psi(x) h(x)-\lambda_1\phi^4(x) -
\lambda_2\phi^2(x) h^2(x) -
\lambda_3 h^4(x)\ ,\nonumber
\ea
with the scalar mass $m^2_h \equiv \left(M^2-2\Delta_2\right)$ and 
the ultra-low energy effective couplings given by 
\ba
&&g_f =
\sqrt{\frac{2\pi^3}{N\Lambda}} 
\int^{+\infty}_{-\infty} dz\ h_J(z)\,\psi^2_0(z) =
\frac{\pi}{4} \sqrt{\frac{M\pi^3}{\Lambda N}}\ ,\no
&&\lambda_1 = \frac{\pi^3}{N\Lambda}
\int^{+\infty}_{-\infty} dz\ \phi_J^4(z) =
\frac{18M\pi^3}{35\Lambda N}\ ,\no
&&\lambda_2 = \frac{\pi^3}{N\Lambda}
\int^{+\infty}_{-\infty} dz\  h_J^2(z)\,\phi_J^2(z)
=\frac{2M\pi^3}{5\Lambda N}\ , \no
&&\lambda_3 = \frac{\pi^3}{N\Lambda}
\int^{+\infty}_{-\infty} dz\ h_J^4(z)
=\frac{M\pi^3}{3\Lambda N}\ .\la{lowi}
\ea 
Herein the ultralow-energy fields $\phi(x)$ and $h(x)$ only have been 
retained, 
whilst the heavy scalars and fermions with masses $\sim M$ have been decoupled.

Quite remarkably, the domain wall Lagrange density \gl{lowi} has a non-trivial
large cut-off limit provided that the ratio
$M/ \Lambda < 1$ is fixed. The four-dimensional ultralow-energy theory
happens to be  interacting with the ratios,
\be
g_f^2:\lambda_1 :\lambda_2 :\lambda_3 \sim 6 : 5 : 4 : 3\ ,
\ee
being independent of high-energy scales and regularization profiles --
here we leave aside the issues concerning the renormalization group improvement.

However the solution $(J)$ is not of our main interest because
the vacuum expectation value of the field $H$ vanishes and does
not supply the domain wall fermion with a light mass.

%%%%%%%%%%%%%%%%%%%%%%%%%%%%%%%%%%%%%%%%%%%%%%%%%%%%%%%%%%%%%%%%%%%%%%%%%%%%%%%%%%%%

\section{Domain walls: Higgs phase}

Evidently, the domain wall solution of Eq.\,\gl{Jvacuum} as well as
the constant background solution of Eq.\,\gl{consol} just
break the $\tau_1$- and $\tau_2$-symmetries of the Lagrange density
\gl{aux}, whereas they keep
the $\tau_3$-invariance untouched -- see Eq.\,\gl{tausim}. 
Meanwhile, the second domain wall
background of Eq.\,\gl{Kvacuum} does break all the $\tau$-symmetries,
{\it i.e.} it realizes a
different phase in which the masses for light particles are naturally created.
We notice however that for the kink solution $(K)$ realizing the space defect
in the $z$-direction the combined parity under the transformations,
\ba
&&z \longrightarrow - z,\quad \Phi(x,-z) \longrightarrow -  \Phi(x,z),\quad 
H(x,-z) \longrightarrow H(x,z), \no
&& \Psi_l(x,-z)  
\longrightarrow \hat\gamma_5 \tau_3  \Psi_l (x,z),
\quad \bar\Psi_l(x,-z)  \longrightarrow - \bar\Psi_l(x,z)\hat\gamma_5 \tau_3, \la{cpar}
\ea
remains unbroken.

As we will see below the mass scale for light particles is controlled 
by the parameter $\mu =\sqrt{2 \Delta_2 - M^2}$, 
which describes the deviation from the critical scaling point
where the two regimes of Eq.s\ \gl{Jvacuum} and \gl{Kvacuum} melt together.

As we want to supply fermions with masses much lower than the threshold of
penetration into the fifth dimension, to protect the four-dimensional physics,
we assume further on that $\mu \ll M$. At the solution $(K)$: 
$S(X) = S_K(z)$,
the $2\times2$ matrix-valued mass operator ${\bf M}[S(X)]$
of Eq.\,\gl{massa} for scalar excitations gets the following entries,
\ba
&&{\rm M}_{K}^{11}=
-\partial_z^2+4M^2 +
2(\mu^2 - 3 M^2)\,{\mbox{\rm sech}^2(\beta z)}\ ,\\
&&{\rm M}_{K}^{22}=
-\partial_z^2+M^2 - \mu^2-2
(M^2-3\mu^2)\,{ \mbox{\rm sech}^2(\beta z)}\ ,\\
&&{\rm M}_{K}^{12}={\rm M}_{K}^{21}=
4 M\mu\,{\mbox{\rm sinh}(\beta z)}\,{ \mbox{\rm sech}^2(\beta z)}\ ,
\ea
the positive quantities $\mu$ and $\beta$ being defined in Eq.\,\gl{mu}.
As this mass operator is non-diagonal it mixes the scalar fields $\Phi$ and
$H$. However, this mixing does fulfill the combined symmetry,
\be
s(x,z) = \pm\,\sigma_3\,s(x, - z)\ ,\quad  {\bf M}[S_K(z)] =
\sigma_3\,{\bf M}[S_K(- z)]\,\sigma_3\ , \la{psym}
\ee
because the diagonal elements of ${\bf M}[S_K(z)]$ are even 
and the off-diagonal
ones are odd with respect to the reflection $z \rightarrow - z$.
This symmetry allows to classify the non-degenerate eigenstates of
the operator ${\bf M}_{K}(z)\equiv {\bf M}\left[S_{K}(z)\right]$ 
according to
their parity. Actually, they may consist of
even-odd  or odd-even pairs of components 
\begin{displaymath}
S(X)-S_K=\left\lgroup\begin{array}{c}
s_1(x,z)\\
s_2(x,z)\end{array}\right\rgroup\ .
\end{displaymath}
One more symmetry can be revealed in respect to the simultaneous change in
the sign of $\mu$ and the reflection $z \rightarrow - z$, that means
\be
{\bf M}[S_K(z,\mu)] =
{\bf M}[S_K(- z,-\mu)]\ . \la{musym}
\ee
This formal symmetry turns out to be crucial in perturbation theory
to build up eigenstates
of ${\bf M}_{K}(z)$  for
small $\mu \ll M$. Indeed, after a
proper normalization of the eigenstates, the expansion in powers of $\mu$
of the latter ones is constrained as follows: the parity-even components of
$s(x,z)$ have to contain even powers of $\mu$, whilst the
parity-odd ones have to contain odd powers of $\mu$.

As a matter of fact,
for $\mu \ll M$ the operator 
${\bf M}_{K}(z)$
can be suitably decomposed into its diagonal part
${\bf M}_{K}^{(0)}$ with the very same diagonal matrix elements
of Eq.s\ \gl{massII1} and \gl{massII2}, in which
$M$ is replaced by $\beta$ and $\mu=0$,
plus the small perturbation $\Delta{\bf M}_{K}$,
\be
{\bf M}_{K}={\bf M}_{K}^{(0)}+\Delta{\bf M}_{K}\ ,\label{pertu}
\ee
where
\begin{displaymath}
{\bf M}_{K}^{(0)} \equiv
\left\lgroup\begin{array}{cc}
- \partial_z^2 + 4\beta^2 -
6\beta^2\,{\mbox{\rm sech}^2(\beta z)}
 & 0\\
0 & - \partial_z^2 + \beta^2-
2\beta^2\,{\mbox{\rm sech}^2(\beta z)}
\end{array}\right\rgroup\ ,
\end{displaymath}
whereas
\begin{displaymath}
\Delta{\bf M}_{K} \equiv 4 \beta^2\epsilon \left\lgroup\begin{array}{cc}
\epsilon\,{\mbox{\rm tanh}^2(\beta z)}
 & \sqrt{1 + \epsilon^2}\
{\mbox{\rm tanh}(\beta z)}\,{\mbox{\rm sech}(\beta z)}\\
\sqrt{1 + \epsilon^2}\
{\mbox{\rm tanh}(\beta z)}\,{ \mbox{\rm sech}(\beta z)}
& \epsilon\,{\mbox{\rm sech}^2(\beta z)}
\end{array}\right\rgroup\ .
\end{displaymath}
Herein the dimensionless parameter 
$\epsilon = \mu/ \beta \ll 1$ controls the deviation 
from the scaling point where the two regimes $(J)$ and $(K)$ do
coincide and both scalars are massless. One can choose the parameters
$\beta$ and $\epsilon$ as independent ones, characterizing the mass unit and
a scaling-point deviation. Then the initial mass scales $M$ and $\mu$
can be conveniently expressed in terms of the latter ones as
\be
M = \beta \sqrt{1 + \epsilon^2}\ ,\qquad \mu = \beta \epsilon\ .
\ee 

Now, it turns out that the operator
${\bf M}_{K}^{(0)}$ has two zero-modes, as  it
immediately follows from the comparison
with the related operators \gl{massII1} and \gl{massII2} in which
 $M \rightarrow \beta$ and  $M^2 - 2 \Delta_2$ is omitted.
The corresponding massless
eigenfunctions are obviously given by Eq.s\ \gl{ground1} and \gl{ground2} 
up to the replacement  $M \rightarrow \beta$. 
The deviation from the scaling point is therefore fully
generated by the perturbation term $\Delta{\bf M}_{K}$, as it does.

The masses of the lightest localized scalar states can
be  obtained again -- see Eq.\,\gl{ground2} -- from the Schr\"odinger
equation with a matrix-valued potential: namely,
\be
\sum_{j=1,2}{\bf M}_{K}^{ij}\Omega^{j}_{K}(X)=
m^2 \Omega^{i}_{K}(X)\ ,\qquad i = 1,2 .
\la{massIII}
\ee
It is easy to find the massless state as a zero-mode
of the operator in Eq.\,\gl{pertu}.
Indeed, differentiating with respect to $z$ the equations \gl{steq}
of the stationary configuration,
one obtains the normalized zero-mass solution
from the
derivative of the kink-like solution $S_K(z)$, with the
upper signs, in the form
\ba
\Omega^\phi_{K}(X) &=& \phi(x)\left\lgroup\begin{array}{c}
\phi^+_K(z)\\
\phi^-_K(z) \end{array}\right\rgroup\frac{\nu}{\sqrt\Lambda} \ , \no
{\bf \phi}_K (z) &\equiv& \left\lgroup\begin{array}{c}
\phi^+_K(z)\\
\phi^-_K(z) \end{array}\right\rgroup \equiv 
\partial_z S_K(z)\,\sqrt{\frac{3}{\beta^3(4 + 6\epsilon^2)}}\no
&=&\sqrt{\frac{3\beta}{4 + 6\epsilon^2}}\,
\left\lgroup\begin{array}{c} \sqrt{1 + \epsilon^2}\
{\mbox{\rm sech}^2 (\beta z)}
\\ - \epsilon\ {\mbox{\rm sinh} (\beta z)}\,
{\mbox{\rm sech}^2 (\beta z)} 
\end{array}\right\rgroup\
\ ,\la{state1}
\ea
according to the translation symmetry breaking in
the kink background \ci{Rajar}.  
Evidently, this eigenfunction is related to the unperturbed
zero-mode of Eq.\,\gl{ground1} at $\mu = 0$ which is parity-even.
Respectively, it consists in turn of the parity-even
upper component $\phi^+_K(z)$ and the parity-odd lower component
$\phi^-_K(z)$,
in accordance to the combined symmetry of Eq.\,\gl{psym}, because
the perturbation 
term does not change the parity properties of an eigenstate.  
Finally, one can convince oneself that, for a given $\beta$,
the expansion in
$\epsilon \leftrightarrow \mu$
contains only even powers of this parameter for
the upper, even component and only its odd powers for the lower, odd 
component
in accordance  with the $\mu$-symmetry of Eq.\,\gl{musym}.

The second light scalar state $\Omega^h_{K}(X)$ arises from 
the perturbation
of the zero-mass state of Eq.\,\gl{ground2} for $\mu = 0$. 
Therefore, it contains
a parity-odd upper component  $h^-_K(z)$ that can be expanded
into odd powers
of $\epsilon$, together with a parity-even lower component  $h^+_K(z)$
which can be expanded in even powers
of $\epsilon$, in such a way that the exact normalized second light scalar
state can be written as
\be
\Omega^h_{K}(X) \equiv h(x)\left\lgroup\begin{array}{c}
h^-_K(z)\\
h^+_K(z) \end{array}\right\rgroup\frac{\nu}{\sqrt\Lambda}\ . \la{state2}
\ee
To the first order in the $\epsilon$ expansion -- see Appendix
C -- the localization functions read
\be
 h_K (z) \equiv \left\lgroup\begin{array}{c}
h^-_K (z)\\
h^+_K(z) \end{array}\right\rgroup
=
\sqrt{\frac{\beta}{2}}
\left\lgroup\begin{array}{c} 
- \epsilon\beta z\ {\mbox{\rm sech}^2 (\beta z)}
\\ 
{\mbox{\rm sech} (\beta z)}  
\end{array}\right\rgroup\ + {\cal O}(\epsilon^2)\ .\la{state22}
\ee
The mass of this scalar state,
to the first order in $\epsilon$ expansion, is given by
\be
m_h = \beta\,\epsilon\sqrt2 + {\cal O}(\epsilon^2)\ \simeq
\mu \sqrt2\ .
\la{massh}
\ee
One can see that now the mass eigenstates enter both in the upper component
$s_1 (X) = \Phi(X) - \langle\Phi(X)\rangle_0$ 
and in the lower component  of
$s_2 (X)= H(X) - \langle H(X)\rangle_0$ 
of the low-energy scalar field: namely,
\ba
&&s_1 (X) = \Bigl[\,\phi(x)\,\phi^+_K(z) + h (x)\,h^-_K(z)\,\Bigr]
\frac{\nu}{\sqrt\Lambda}\ ,\no
&&s_2 (X) =  \Bigl[\,\phi(x)\,\phi^-_K(z) + h (x)\,h^+_K(z)\,\Bigr]
\frac{\nu}{\sqrt\Lambda}\ . \la{mixt}
\ea
However, the admixture of the opposite-parity states 
is strongly suppressed in the ultralow-energy effective Lagrange density
in the Minkowski space-time, owing to the integration over the
extra-dimension and to the high order in $\epsilon$.

As well as for the solution $(J)$, in the
vicinity of the vacuum solution $(K)$ of Eq.\,\gl{Kvacuum},
the ultralow-energy effective Lagrange density for the light
states can be obtained from
Eq.\,\gl{lowmin} in a similar form. In the leading and next-to-leading 
order of $\epsilon$-expansion,
one can show that the only difference concerns the appearance
of the mass for light fermions and of a cubic
scalar interaction that we calculate for positive values of
$\langle \Phi(X)\rangle_0$ and $\langle H(X)\rangle_0$ in 
Eq.~\gl{Kvacuum}: namely,
\be
{\cal L}_{K}^{(4)}\left[\,\overline{\psi},\psi,\phi,h\,\right] =
{\cal L}_{J}^{(4)}\left[\,\overline{\psi},\psi,\phi,h\,\right]
 - m_f\,\overline{\psi}(x)\psi(x)
-  \lambda_4\,h^3(x)+ {\cal O}(\epsilon^2)\ , \la{lagr4K}
\ee
the fermion mass being determined by the expression
\be
m_f \equiv \int^{+\infty}_{-\infty} dz\,
\overline{\psi}_0(z)\, H_K (z)\,\psi_0(z)
= \frac{\pi}{4}\,\beta\,\epsilon  =   \frac{\pi}{4}\,\mu \ ,
\la{fermas}
\ee
where Eq.s~\gl{Kvacuum} and \gl{psi0} have been employed, whereas
\begin{eqnarray}
\lambda_4 &=& \sqrt{\frac{8\pi^3}{\Lambda N}}\int^{+\infty}_{-\infty}
dz\,\Bigl[h^+_K(z)\Bigr]^2
\left\{ \Phi_K (z)\,
h^-_K(z) +
H_K (z)\,h^+_K(z)\right\}\no
 &\approx& \mu\pi\,\sqrt{\frac{M \pi}{\Lambda N}}\ .\la{constf}
\end{eqnarray}
We stress that a generally {\it a priori}
possible 3-point vertex $\phi^2(x)\,h(x)$ does actually
receive mutually compensating contributions from different mixing terms.
Thereby the direct decay of the massive Higgs-like boson $h$ into a pair of
massless branons \ci{bran3} is suppressed and the low-energy
Standard Model matter turns out to be stable.

To sum up, in the present model the ratio of the dynamical fermion mass 
(presumably the top quark one \ci{Topmode}) 
to the Higgs-like scalar mass is firmly 
predicted
to be
\be
m_h : m_f = 4\sqrt{2}: \pi \simeq 1.8\ ,
\ee
which is somewhat less than such a ratio as predicted in the Top-mode
Standard Model -- the Nambu relation gives $m_h : m_f = 2$.

Finally, concerning
the coupling parameters $g_f^2,\, \lambda_1,\,\lambda_2,\, \lambda_3$,
they are
essentially described by Eq.s~\gl{lowi}, up to the
two orders in $\epsilon$-expansion, the corrections being of
the order ${\cal O}(\epsilon^2)$.
We would like to stress that the coupling of fermions 
to the massless scalar
$\phi(x)$ does no longer appear even in the next order $\sim \epsilon^2$,
because its mixing
form factor $\phi^-_K(z)$ in Eq.\,\gl{mixt} is an odd function,  making
thereby the  relevant integral of Eq.\, \gl{lowi} identically vanishing.
\\
%%%%%%%%%%%%%%%%%%%%%%%%%%%%%%%%%%%%%%%%%%%%%%%%%%%%%%%%%%%%%%%%%%%%%%%%%%%%%%%%%

\section{Manifest breaking of translational invariance}

One can conceive that in reality the translational invariance in 
five dimensions is broken not only spontaneously but also manifestly 
due to the presence of a gravitational background, of other branes etc.
In a full analogy with the Goldstone boson physics one can expect \ci{bran3} 
that the manifest breaking of translational symmetry is such to supply
the branons with a mass.  

In the model presented in our paper the natural realization of the
translational symmetry breaking can be implemented 
by the inhomogeneous scalar 
background fields coupling to  the lowest-dimensional fermion currents.
Let us restrict ourselves to the scenario of the type $(K)$
and introduce two scalar
field defects with the help of the functions
${\rm F}_\Phi(z)$ and ${\rm F}_H(z)$, which are supposed to be quite
small, {\it i.e.}, $|{\rm F}_{\Phi}(z)|\ll |\langle\Phi(X)\rangle_0|$ and
$|{\rm F}_{H}(z)|\ll |\langle H(X)\rangle_0|$. These scalar defect fields
catalyze the translational
symmetry breaking and the domain wall formation by means of their
interactions with the fermion currents,
\be
{\cal L}^{(5)}_{\rm F} =\ -\ {\rm F}_\Phi(z)\,
\overline{\Psi}(X)\tau_3\Psi(X)\ -\ {\rm F}_H(z)\,
\overline{\Psi}(X)\tau_1\Psi(X)\ . \la{lagdef}
\ee
When supplementing the Lagrangian \gl{aux}
under the Hubbard-Stratonovich transformation, one can see
that those background defect fields do actually couple to the auxiliary 
scalar fields in Eq.~\gl{aux} after the replacements
\be
 \Phi\ \longmapsto\ \Phi - {\rm F}_\Phi(z)\ ,
\qquad 
H\ \longmapsto\  H - {\rm F}_H(z)\ .
\ee
The latter redefinition  of the auxiliary fields reveals the explicit
coupling of defect fields to the auxiliary scalars, which dictates in turn
the eventual change of the scalar field Lagrange density \gl{lowmin}: namely,
\ba
&&{\cal L}^{(5)}\left[\,\overline{\Psi},
\Psi,\Phi, {\rm F}_\Phi(z), H, {\rm F}_H(z)\,\right] =
{\cal L}^{(5)} (\overline{\Psi},\Psi,\Phi, H)+ \no
&&\frac{2N\Lambda^3}{g_1}\,\Phi\,{\rm F}_\Phi(z)  +
\frac{2N\Lambda^3}{g_2}\,H\,{\rm F}_H(z)
- \frac{N\Lambda^3}{g_1}\,{\rm F}_\Phi^2(z) -
\frac{N\Lambda^3}{g_2}\,{\rm F}_H^2(z)\ . \la{lowd}
\ea
In order to prepare the scalar part of the low-energy effective 
action in comparable units let us re-scale,
\be
{\rm F}_\Phi(z) \equiv
\frac{g_1\mu^3}{4\pi^3\Lambda^2}\,{\rm f}_\Phi(z)\ ,\qquad\quad
{\rm F}_H(z) \equiv
\frac{g_2\mu^3}{4\pi^3\Lambda^2}\,{\rm f}_H(z)\ , \la{resc}
\ee
where we have kept in mind that $g_{1} \simeq g_{2} \simeq g^{\rm cr} = 9\pi^3$.
If we assume the dimensionless functions ${\rm f}_{\Phi}$
and  ${\rm f}_{H}$ to be  ${\cal O}(1)$, then
the last two terms in
Eq.~\gl{lowd} are of the order $\mu/\Lambda$ and thereby negligible within the
approximation adopted in this paper.

Taking the notations of Eq.~\gl{mu} into account,
it is not difficult to show that the effective
Lagrange density for low-energy scalar fields in the 
presence of defects can be cast in the form
\ba
&&{\cal L}^{(5)}_{\rm scalar}\left[\,\Phi, H, {\rm f}_\Phi(z),
{\rm f}_H(z)\,\right]  =
\frac{N\Lambda}{4\pi^3}\,
\left[\,\ 2 \mu^3\,{\rm f}_\Phi(z)\,\Phi
+ 2\mu^3\,{\rm f}_H(z)\,H\right. \no
&&\left. +\ (\partial_\alpha \Phi)^2 +
(\partial_\alpha H)^2 + 2 M^2 \Phi^2 + (M^2 + \mu^2) H^2
- (\Phi^2 + H^2)^2\,\right]\ .\la{slagd}
\ea 
The stationary vacuum configurations of scalar fields obey now the
modified set of Eq.~\gl{steq}: namely,
\ba
&&\partial^\alpha\partial_\alpha\Phi = \mu^3 {\rm f}_\Phi(z) +
2\Phi\left[\,M^2 - \Phi^2 - H^2\,\right]\ ,\no
&& \partial^\alpha\partial_\alpha H = \mu^3 {\rm f}_H(z) +
H\,\left[\,M^2 + \mu^2 - 2H^2 - 2 \Phi^2\,\right]\ .
\la{steqd}
\ea
Let us search for the solutions of these equations
generating a domain wall of type $(K)$ which preserves the symmetry \gl{cpar}.
Evidently, the latter ones can be actually
achieved if both the scalar defect fields and the
very vacuum solutions have the same definite parity properties, {\it i.e.},
${\rm f}_\Phi(z)$ and $\Phi_K(z)$
being odd functions of $z$, while
${\rm f}_H (z)$ and $H_K(z)$ being even ones.
Thus we tune ourselves to what we could call 
the collinear mechanism of a brane creation.
As we shall see here
below, it allows to formulate within the perturbation theory
the {\sl self-consistent catalyzation} of a domain wall
in the presence of some weak defects, {\it i.e.},
for sufficiently small defect functions ${\rm f}_{\Phi}(z)$ and
${\rm f}_H(z)$.

Indeed, let us accept the latter assumption and write down the trial
solutions as the sum of the
unperturbed functions $\Phi_K(z)\equiv M\mbox{\rm tanh}(\beta z),\,
H_K (z)\equiv \mu\mbox{\rm sech}(\beta z)$ given in Eq.s~\gl{Kvacuum} and
the corresponding small deviations $\delta_{\rm f}\Phi(z),\delta_{\rm f}H(z)$ 
due to the presence of a weak defect, so that
\be
S_{\rm f} (z) = \left\lgroup\begin{array}{c}
\Phi_{\rm f} (z) \\ H_({\rm f}z)
\end{array}\right\rgroup = \left\lgroup\begin{array}{c}
\Phi_K (z)+ \delta_{\rm f}\Phi(z) \\ H_K(z) +\delta_{\rm f} H(z)
\end{array}\right\rgroup \equiv S_K(z) +  \delta_{\rm f} S (z)\ .
\ee
To the first order in deviation functions, Eq.s~\gl{steqd} entail
\be
{\bf M}\left[S_K(z)\right]\,\delta_{\rm f} S\equiv
{\bf M}_K(z)\,\delta_{\rm f} S =
\mu^3 {\rm f}\ ,\quad {\rm f}(z) \equiv
\left\lgroup\begin{array}{c}
{\rm f}_\Phi(z) \\ {\rm f}_H (z)
\end{array}\right\rgroup\ ,\label{deltaf}
\ee
where the second variation operator ${\bf M}_{K}(z)$, already
introduced in Eq.~\gl{pertu}, does appear in this inhomogeneous equation.
Its solution can be suitably presented by means of the spectral decomposition
for the operator ${\bf M}_{K}(z)$ itself. In fact,
as it was elucidated in the preceding Section, its spectral
resolution contains
two sets of eigenfunctions with opposite parity properties. In particular,
once we search for solutions of odd upper component and even 
lower component type, the relevant part of the spectral decomposition
of the operator ${\bf M}_{K}(z)$ consists of the projector onto
the bound state,
Eq.~\gl{state22}, belonging to the eigenvalue $m_h^2\simeq 2\mu^2$,
together with the projectors onto the continuum improper states,
starting from the eigenvalue $M^2 \gg 2\mu^2$: namely,
\be
{\bf M}_K(z)=m_h^2\,{\widehat{\bf P}}_h+
\int_{M^2}^\infty d\sigma\,\varrho(\sigma)
\sigma\,{\widehat{\bf P}}_\sigma+\ {\rm irrelevant}\ ,\la{specdec}
\ee
where ${\widehat{\bf P}}_h,{\widehat{\bf P}}_\sigma$ 
are the projectors onto the proper
and improper states with the relevant parity properties respectively, whereas
$\sigma$ denotes the invariant mass of the continuous part of the spectrum and
$\varrho(\sigma)$ the corresponding density of the states. After inversion 
of the operator \gl{specdec} the approximate solution of 
Eq.~(\ref{deltaf}) reads
\be
\delta_{\rm f}S\simeq \mu\left\{\frac{1}{2}\,{\widehat{\bf P}}_h{\rm f}
+\int_{M^2}^\infty d\sigma\,\varrho(\sigma)\,
\frac{\mu^2}{\sigma}{\widehat{\bf P}}_\sigma{\rm f}\right\}\ .
\ee
Thus, taking into account that $\mu^2\ll M^2 <\sigma$, a finite
solution $\sim \mu $ exists iff the
projection of the defect function ${\rm f}(z)$ onto the bound state is not
vanishingly small. Then the first order approximate solution is given by
\be
\delta_{\rm f}S (z) = \xi\,\mu\,\sqrt{\frac{2}{\beta}}\, h_K (z)
+ {\cal O}\left(\mu \epsilon^2, \mu\xi^2\right)\ ,\la{domin}
\ee
where the dimensionless parameter $\xi$ is introduced by means of the relation,
\be
\xi \equiv\ \sqrt{\frac{\beta}{8}}\,\int_{-\infty}^{+\infty} dz\
\Bigl[\,h^-_K(z)\,
{\rm f}_\Phi(z)
+  h^+_K(z)\,{\rm f}_H (z)\,\Bigr]\ . \la{defxi}
\ee
Let us first assume that  $\epsilon\ll |\xi| \ll1$.
Then, for a wide variety of scalar defects ${\rm f}(z)$, 
the dynamical mechanism of 
the domain wall formation is essentially triggered by the light bound state
of the mass operator ${\bf M}_{K}$, up to the first order approximation in
perturbation theory.
Hence, for this variety, the manifest breaking of translational invariance is
labeled by a one-parameter family of weak defect amplitudes\quad 
$|\xi|\le 1$.
However,
it turns out that in the calculation of the branon mass the
next-to-leading terms -- quadratic in $\xi$ -- are indeed required.
The latter ones are more sensitive to the high-frequency modes
in the spectrum of
the operator ${\bf M}_{K}$, owing to the non-linear interaction 
in Eq.s~\gl{steqd}.
Consequently, they appear to be dependent upon the details of
the defect function ${\rm f} (z)$.
The analysis concerning the general solution will not be addressed here,
instead of,  let us make the {\it ansatz} for
${\rm f} (z)$ which allows to solve Eq.s~\gl{steqd} exactly.
Firstly, one is evidently  free to restrict
the allowed set of defect functions in such a way that the solution
\gl{domin} keeps itself reliable all over the range $\epsilon\ll|\xi|<1$.
In particular, it is provided by the \underline{ansatz $(A)$}\ :
\ba
&&\mu^3 {\rm f}^{(A)}_\Phi\ \simeq\ \xi\,\mu^3 \sqrt{\frac{8}{\beta}}
\,h^-_K  +\no
&& 6 \Phi_K(\delta_{\rm f}\Phi)^2  +
4 H_K\,\delta_{\rm f} H\,\delta_{\rm f}\Phi
+ 2 (\delta_{\rm f}\Phi)^3 + 2 (\delta_{\rm f} H)^2 \delta_{\rm f}\Phi\ ;\no
&&\mu^3 {\rm f}^{(A)}_H\ \simeq\ \xi\,\mu^3 \sqrt{\frac{8}{\beta}}
\,h^+_K  + \no
&&6  H_K (\delta_{\rm f} H)^2 +
4 \Phi_K\,\delta_{\rm f}\Phi\,\delta_{\rm f} H + 
2 (\delta_{\rm f} H)^3 + 2 (\delta_{\rm f}\Phi)^2 \delta_{\rm f} H\ ,
\la{barf}
\ea
where $ \Phi_K, H_K $ are given by Eq.s~\gl{Kvacuum} and the  
deviations 
$\delta_{\rm f} \Phi, \delta_{\rm f} H$ are 
taken from Eq.~\gl{domin}.
For such a defect, to all orders in the weak defect amplitude,
the modified kink configuration of the scalar fields reads
\be
S_K^{(A)}(z) \approx \beta \left\lgroup\begin{array}{c}
   \left(1 + \frac12\epsilon^2\right)\mbox{\rm tanh}(\beta z) - \xi
\epsilon^2 \beta z\ {\mbox{\rm sech}^2 (\beta z)}\\
\epsilon (1+ \xi) \,\mbox{\rm sech}(\beta z) 
\end{array}\right\rgroup. \la{linsol}
\ee
As the lightest domain wall fermion gets its mass only from the lower component
of  $S_K^{(A)} (z)$, then the mass correction is just reproduced
by the following
multiplicative factor of
the unperturbed mass \gl{fermas},
\be
m_f \approx  \frac{\pi}{4}\,\mu\  (1+ \xi) \ .\la{fermshift}
\ee
In terms of
$\beta$ and $\epsilon$, the mass operator for scalar particles receives an
additional perturbation
as compared to Eq.s~\gl{pertu}: namely,
\ba
&&{\bf M}_K^{(A)}= {\bf M}_{K}^{(0)}+\Delta{\bf M}_{K}
+ \Delta{\bf M}_A\ ,\label{defmA}\\
&&\Delta{\bf M}_A
\equiv 4\beta^2\epsilon\,\xi\, {\mbox{\rm sech}^2(\beta z)} \no
&&\times\left\lgroup\begin{array}{cc}
\epsilon\,
\left[\,1+ \frac12 \xi\ - 3 \beta z\ {\mbox{\rm tanh}(\beta z)}\,\right]
 &
{\mbox{\rm sinh}(\beta z)} \\
{\mbox{\rm sinh}(\beta z)}
& \epsilon\,\left[\,
3+ \frac32 \xi -  \beta z\ {\mbox{\rm tanh}(\beta z)}\,\right]
\end{array}\right\rgroup \no
&&\times \ \left[\,1 + {\cal O} \left(\epsilon^2\right)\,\right]\ , \nonumber
\ea 
where the basic definition of the second variation operator 
Eq.~\gl{massa} has been used.
Therefrom one obtains the masses of scalar states -- see Appendix D.
In particular, the lightest branon is no longer massless
\be
\left(m^{(A)}_\phi\right)^2
\approx \frac45 \mu^2\, \xi^2
\ee
and its localization function turns out to be
\be
{\bf \phi}_A (z)  \approx
\sqrt{\frac{3\beta}{4}}\,
\left\lgroup\begin{array}{c} 
{\mbox{\rm sech}^2 (\beta z)}
\\ - \epsilon\ (1 + \xi)\ {\mbox{\rm sinh} (\beta z)}\,
{\mbox{\rm sech}^2 (\beta z)} 
\end{array}\right\rgroup
\ .\la{statef1}
\ee
Respectively, the correction to the Higgs boson mass reads,
\be
\left(m^{(A)}_h\right)^2 \approx
2 \mu^2\left(1 + 3\xi +\frac53 \xi^2\right)\ ,\la{higA}
\ee
whilst the corresponding localization function is given by
\be
 h_A (z) \approx
\sqrt{\frac{\beta}{2}}
\left\lgroup\begin{array}{c} 
- \epsilon \, (1 + \xi)\ \beta z\, {\mbox{\rm sech}^2 (\beta z)}
\\ 
{\mbox{\rm sech} (\beta z)}  
\end{array}\right\rgroup\ .\la{statef2}
\ee

We see that the regular defect does not rigidly polarize the domain wall
vacuum $(K)$ in a fixed direction  -- the sign of  $\xi$ can be not only
positive but also negative while delivering a local minimum of the induced
scalar field Hamiltonian, {\it i.e.} physical masses for scalar bosons.
 For the latter case, $\xi < 0$, 
one can reduce the mass of
Higgs particle as compared to the fermion (`` top-quark'') mass,
\be
\left(m^{(A)}_h\right)^2 : m^2_f =
\frac{32 (1 + 3\xi +5\xi^2/3)}{\pi^2 (1 +\xi)^2}\ .
\ee
For instance, if $\xi \simeq - 0.4$ and the fermion mass is assumed to be
of order of the top-quark mass $\sim$ 175 GeV, then the Higgs mass is estimated
to be $\sim$ 135 GeV, which is acceptable from the phenomenological
viewpoint \ci{PDG}. For the same choice the branon mass is found to be
$\sim$ 100 GeV. However, we stress that the predictions for scalar masses are
essentially based on contributions which are
quadratic in $\xi$, especially for the branon
mass. But the latter terms depend on a model for the
space defect. 

Both the localization functions \gl{statef1}, \gl{statef2}  
contain corrections linear in $\xi$ which
are the only ones relevant for calculation of the induced 3-point and
4-point vertices for low-energy scalar fields. The coupling constants 
$g^2_f, \lambda_1,\lambda_2,\lambda_3 $ in the 4-point vertices
on the domain wall are built of the components $\phi^+_A, h^+_A$, which are
universally given by Eq.s~\gl{lowi} at the leading order in $\epsilon^2$
and thereby do not depend on a small background defect. The coupling constant 
$\lambda_4$ in the 3-point vertex $h^3(x)$
is described by the integral in Eq.~\gl{constf}
which is evidently homogeneous in the multiplier $(1 + \xi)$. Hence, in
the presence of a defect,  this coupling is simply renormalized as follows,
\be
 \lambda_4^{({\rm f})} \approx  (1 + \xi)  \lambda_4 \approx 
(1 + \xi)\mu\pi\,\sqrt{\frac{M \pi}{\Lambda N}} \ .
\la{constfA}
\ee
Another possible induced coupling constant in the vertex $\phi^2(x)\,h(x)$ 
is also homogeneous in the multiplier $(1 + \xi)$. Thus it remains
vanishing, the Higgs boson does not decay into branons and the 
Standard Model matter is stable.

The previous ansatz $(A)$ has been a representative of non-topological or
regular defects with background functions decreasing at the infinity. 
Now we shall introduce and discuss another kind of defect of 
a topological type, the influence of which on the
branon masses is drastically different.
Let us take the following \underline{ansatz $(B)$} for the defect ${\rm f}(z)$
providing the analytical solution for vacuum configurations,
\be
 \mu\,{\rm f}^{(B)}(z) =\left\lgroup\begin{array}{c}
M \,\gamma\, \mbox{\rm tanh}(\bar\beta z) \\ \mu\, \kappa\, 
\mbox{\rm sech}(\bar\beta z)
\end{array}\right\rgroup\ ,
\ee 
where $\gamma, \kappa$ are dimensionless parameters. Notice that the
components are just borrowed from the unperturbed solution \gl{Kvacuum}.
We accept that the upper
component can be  enhanced including the normalization on $M$ rather 
than on $\mu$ -- otherwise it becomes irrelevant as it will be 
clear later on.
We also remark that, on the one hand, the related function is not square
integrable and lies outside the domain spanned by the normalizable
eigenfunctions of the operator ${\bf M}_{K}$. On the other hand, it
represents a genuine topological defect whereas the lower, square
integrable component does not influence the field topology,
{\it i.e.}, it is a regular defect.

The exact solution of Eq.s~\gl{steqd} keeps the form of Eq.~\gl{Kvacuum},
although its coefficients are now controlled by $\gamma, \kappa$: namely,
\be
S_K^{(B)} (z) = \left\lgroup\begin{array}{c}
   M \,a\, \mbox{\rm tanh}(\bar\beta z) \\
\mu \,(1 + \xi ) \,\mbox{\rm sech}(\bar \beta z)
\end{array}\right\rgroup\ . \la{solans}
\ee 
The constants $a, \xi, \bar\beta$ obey the relations,
\ba
&& 2(a^2 - 1) a = \frac{\gamma\epsilon^2}{1 + \epsilon^2}\ ,
\qquad a \approx 1 + \frac14  \gamma \epsilon^2\ ,\\
&& \left(2\xi + \xi^2 + \frac{\gamma}{2a}\right)(1 + \xi ) = 
\kappa\ , \la{kappa} \\
&&\bar\beta^2 = M^2 a^2  - \mu^2  (1 + \xi )^2 \approx
\beta^2 \left[\, 1 + \epsilon^2 \left(\frac12\gamma - 2\xi 
- \xi^2\right)\,\right]\ .
\ea
From the last equation one can reveal a certain 
interplay between the enhanced topological defect and
the regular defect. Instead of the inversion of Eq.~\gl{kappa} 
we prefer to use
$ \xi$ as an independent parameter, 
thereby applying this relation as a definition of $\kappa$.

Let us expand now the solution \gl{solans} in powers of $\epsilon$,
\ba
&&S_K^{(B)} (z) \approx \la{ans2}\\ 
&&\beta \left\lgroup\begin{array}{c}
   \Bigl[\,1 +\frac14\epsilon^2(2 + \gamma)\,\Bigr]\mbox{\rm tanh}(\beta z)
+ \frac14
\epsilon^2 (\gamma - 4\xi - 2\xi^2) \beta z\ {\mbox{\rm sech}^2 (\beta z)}
\\
\epsilon (1+ \xi) \,\mbox{\rm sech}(\beta z) 
 \,
\end{array}\right\rgroup. \nonumber
\ea
Here one can see that the linear analysis in Eq.~\gl{domin} is again valid
and entails $\xi = (\kappa/2) - (\gamma/4)$ in accordance with
the linear part of  the exact solution \gl{kappa}. As the lower, even component
of Eq.~\gl{ans2} coincides with that one in Eq.~\gl{linsol},
the shift in the fermion
mass remains the same as in Eq.~\gl{fermshift}.

The mass operator for this ansatz $(B)$ gets a similar
functional structure albeit
different coefficients: namely,
\ba
&&{\bf M}_K^{(B)}= {\bf M}_{K}^{(0)}+\Delta{\bf M}_{K}
+ \Delta{\bf M}_B\ ,\no
&& \Delta M_B^{11} \equiv \beta^2\epsilon^2\,{\mbox{\rm sech}^2(\beta z)} 
\left[4 \xi\  + 2 \xi^2 \right.\no
&&\left.
+ 3 (\gamma - 4 \xi\  - 2 \xi^2)\,
\beta z \,{\mbox{\rm tanh} (\beta z)}+ 
3 \gamma {\mbox{\rm sinh}^2 (\beta z)}\right]\ 
+ {\cal O} \left(\epsilon^4\right)\ ,\no
&& \Delta M_B^{12} \equiv \Delta M_B^{21} =  4 \xi \beta^2 
\epsilon\,{\mbox{\rm sech} (\beta z)}
{\mbox{\rm tanh}(\beta z)}
+ {\cal O} \left(\epsilon^3\right)\, , \no
&&\Delta M_B^{22} \equiv \beta^2\epsilon^2\,{\mbox{\rm sech}^2(\beta z)} 
\left[12 \xi\  + 6 \xi^2 
\right.\no
&&\left.+  (\gamma - 4 \xi\  - 2 \xi^2)\,
\beta z \,{\mbox{\rm tanh} (\beta z)} 
+  \gamma {\mbox{\rm sinh}^2 (\beta z)}\right]\ 
+ {\cal O} \left(\epsilon^4\right)\  . \label{defmB}
\ea 
The corresponding mass spectrum of lightest scalar states is calculated in
Appendix D. In particular, 
the branon mass is no longer governed by terms quadratic in $\xi$,
\be
\left(m^{(B)}_\phi\right)^2
\approx \gamma \mu^2\ .
\ee
Now, the striking evidence occurs for a strong polarization effect induced by  
a topological defect: the local minimum is guaranteed only
for asymptotics at infinities which are coherent in their signs, {\it i.e.}, 
for positive $\gamma$.

The Higgs mass encodes both topological and non-topological vacuum
perturbations,
\be
\left(m^{(B)}_h\right)^2 \approx \mu^2\,\left(2 + 6\xi + 
3 \xi^2 +\frac12 \gamma\right)\ .
\la{higB}
\ee 
As $\gamma > 0$ the topological defect makes the Higgs particle heavier. 
However, once again, the sign of $\xi$ is not fixed by the requirement 
to provide
a local minimum. Therefore one can find a window for a relatively light Higgs
scalar. If branons are very light, {\it viz.} $\xi \gg \gamma$, then
one can neglect the
last term in Eq.~\gl{higB} and reproduce the Higgs mass \gl{higA} 
of ansatz $(A)$ with a reasonable precision. 

As to  the localization functions for scalar states, they coincide with
those ones of ansatz $(A)$. Thus they are universally parameterized by
terms linear in $\xi$, to the leading orders in $\epsilon$-expansion.
Respectively, the induced coupling constants 
$g^2_f, \lambda_1,\lambda_2,\lambda_3, \lambda_4$
appear to be universal as well.

%%%%%%%%%%%%%%%%%%%%%%%%%%%%%%%%%%%%%%%%%%%%%%%%%%%%%%%%%%%%%%%%%%%%%%%%%%%%%%%%%%%%%

\section{Conclusions and discussion}

The main task of the present work has been the explanation 
of how light fermions
and Higgs bosons may be located on the four-dimensional domain wall 
representing our Universe. The five-dimensional fermions with
their strong self-interactions have been used to discover
the vacuum that breaks spontaneously the translation invariance
at low energies. Its remarkable
feature is the appearance of a topological defect
which induces the light particle trapping. 
Depending on the relation of the
four-fermion coupling constants, the light fermion
states may remain massless or achieve
the masses $\sim \mu$, supposedly, 
much less than the characteristic scale $M$ of the
dynamical symmetry breaking.
These two phases differ in the level of
symmetry breaking. Namely, the internal discrete $\tau$-symmetry of the
initial fermion Lagrange density is generated by the algebra of Pauli matrices
and this symmetry can be broken partially or completely. 
In the latter case,
the light fermions acquire a dynamical mass which is assumed to be
$\sim \mu \ll M$ and
the ratio of the Higgs boson mass to this fermion mass is $\sim 1.8$ close 
to the ratio in the Top-mode Standard Model \ci{Topmode}.

As a consequence of spontaneous breaking of the 
$\tau$-symmetry and of translational invariance two composite scalar bosons
arise, one of which is certainly a Higgs-like massive state whilst
the other one is a Goldstone-like state -- 
a massless domain-wall excitation called branon
\ci{bran2,bran3}. On the one hand, the latter particles turn out to be 
sterile because in the ultra-low
energy world they do not couple directly to the light fermions. On the other
hand, their original coupling to fermions in Eq.~\gl{lowmin} mixes light
fermions and 
the ultra-heavy fermion states with masses $\sim M$. Evidently, the amplitude 
of branon production in the light fermion annihilation 
due to exchange by an ultra-heavy
fermion state $ \sim g_f^2\ m^2_f / M^2$ (see Fig.~1) 
is strongly suppressed if the
domain wall thickness $ \sim 1/M$ 
is much less than the Compton length for light particles,
$ 1/M \ll 1/m_f$.    

\FIGURE[p]{\epsfig{file=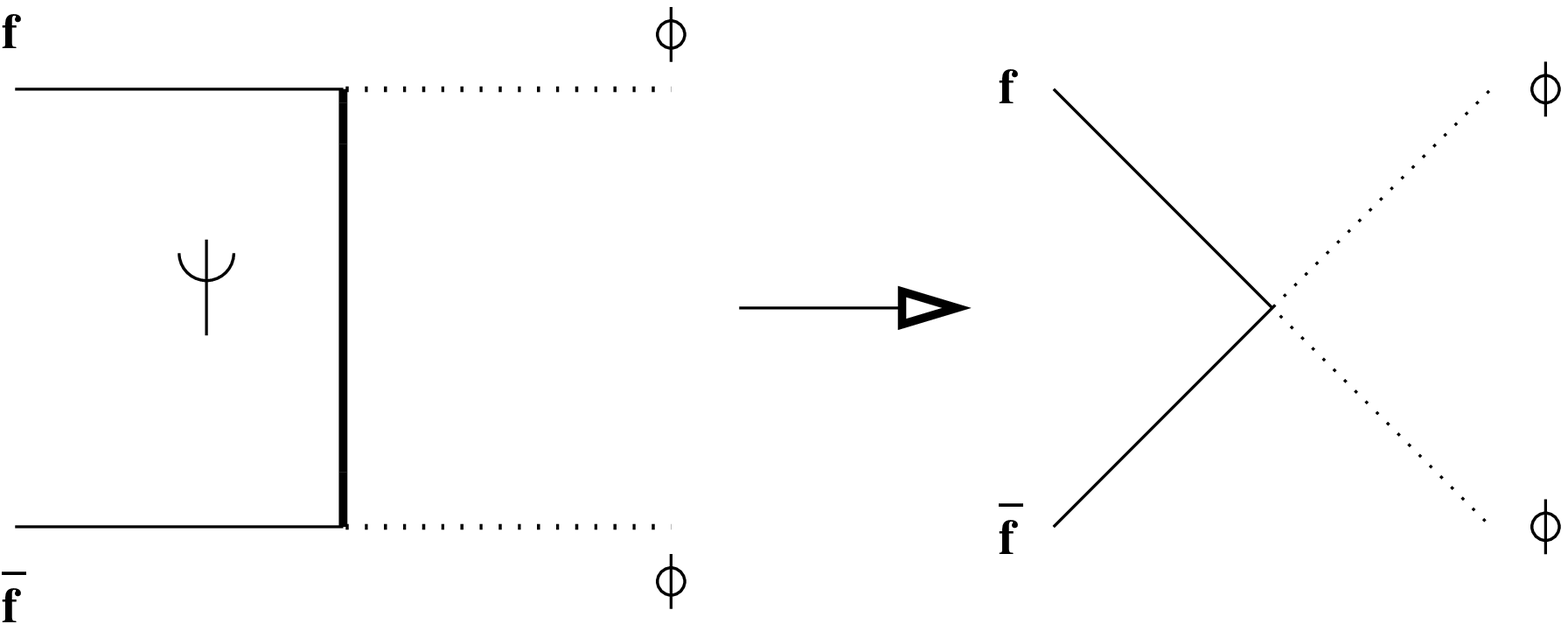,height=4cm,width=7cm}\caption{
Branon $\phi$ creation from light
fermions $f, \bar f$ due to super-heavy fermion $\Psi$ exchange.}}

However, it is quite plausible that the translation and $\tau$-symmetries are
manifestly broken just because the five-dimensional world contains a classical
background of gravitational and various matter fields with a net result
similar to a crystal defect. In the present paper we
have assumed the scalar nature of a five-dimensional space-time defect, which
triggers the domain-wall formation through its coupling to the low-dimensional
fermion operators. The two types of such defects have been investigated.

The first type is a non-topological one, the shape of which belongs to the
Hilbert space of square integrable functions. In this case it happens that 
the only relevant dimensionless parameter $\xi$
governing the manifest symmetry breaking is  given by a  projection
of the defect profile function on the Higgs boson wave function. 
The occurrence of such a defect
supplies the branons with a small mass $\sim \xi\mu$ though its numerical
value is rather model dependent.
Meantime, the polarization of the domain wall background, {\it i.e.} its sign, 
is not correlated with the sign of a non-topological defect -- 
for any sign, the local
minimum of the low-energy effective action is attained.

The second type of defects is a topological one, with different non-vanishing
asymptotics at the infinity. It makes much stronger the catalyzation of a 
domain wall, namely, the minimum and respectively the physical branon masses
are
reached only for coherent signs of the profiles of defect and domain-wall.
In the latter case, again, the branons are supplied with a small mass.

Concerning the Higgs boson masses, their ratio to the fermion 
($\sim$ top-quark)
mass can be substantially reduced with an appropriate choice of a 
non-topological part of the defect $\sim \xi$. In particular, the Higgs masses 
may
be well adjusted to a phenomenologically acceptable \ci{PDG} 
value $\sim 135$ GeV for a reasonably small value of
a defect $\xi \sim 0.4$.
  
In all scenarios the ratios of the induced coupling constants 
in the four-dimensional effective action for light particles are predicted
unambiguously and in a model-independent way. Their absolute values are
universally parameterized by the ratio of
the symmetry breaking scale $M$ to the high-energy (compositeness)
scale $\Lambda$ -- a cutoff in the four-fermion interaction. 
This ratio
$M/\Lambda < 1$ can be thought of as an empirical input 
which must be established from the relevant experiments by measuring, 
for instance,  the Yukawa coupling constant $g_f$ in the direct Higgs 
production {\it via} the fusion mechanism.

 \FIGURE[p]{\epsfig{file=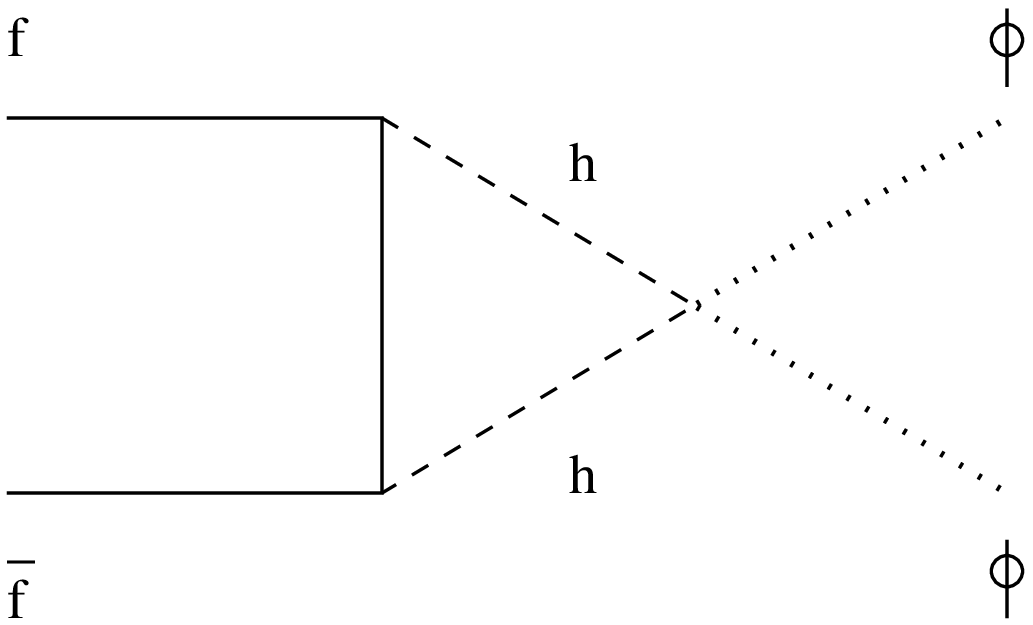,height=4cm,width=6cm}\caption{
Missing energy in fermion
annihilation due
to branon emission.}}

However, the branon matter is much more elusive for experiments because 
it is not produced by fermions at the leading order in the Yukawa coupling 
and does not appear as a decay product of the Higgs bosons.
Still the fermion annihilation (see, Fig.~2)
may result in missing energy events due to radiative creation of branon pair
mediated by Higgs bosons. The corresponding Feynman integral is well 
convergent and therefore, at very low momenta, the amplitude of this process
can be estimated to be of order $g_f^2\cdot \lambda_2\cdot m^2_f / \mu^2$.
We notice that if they are produced by very light fermions, {\it e.g.}
$e^+e^-$, the very last ratio is negligibly small putting such
signals into the experimental background. Nonetheless, for top-quarks (again
in the fusion production) there might be a possibility to discover
branon pair signals for sufficiently light branons. More 
qualitative estimates will be presented elsewhere.
Anyhow, the sterile branons seem to be good candidates for saturation of
the dark matter of our Universe -- see Ref.~\ci{bran3} for a more detailed 
motivation.

\acknowledgments

One of us (A.A.) is grateful to V.A.Rubakov for multiple 
discussions during his visit to the University of Barcelona which
were indeed stimulating to start the present work.
We also acknowledge the long-term correspondence with R. Rodenberg whose
erudition helped us a lot to get orientated in the abundant literature about 
the particle physics from extra dimensions. 
This work is supported by Grant INFN/IS-PI13.
A. A. and V. A. are also supported by Grant RFBR 01-02-17152, 
INTAS Call 2000 Grant (Project 587) and The Program
"Universities of Russia: Fundamental Investigations"\, 
(Grant UR.02.01.001).

%%%%%%%%%%%%%%%%%%%%%%%%%%%%%%%%%%%%%%%%%%%%%%%%%%%%%%%%%%%%%%%%%%%%%%%%%%%%%%%%%%

\appendix
\section{One-loop effective action}
Let us consider the second order matrix-valued positive
elliptic differential operator
\ba
D^\dagger D &=& -\partial_\mu\partial_\mu - \partial_z^2 +
{\cal M}^2(X)=-\partial^2+{\cal M}^2(X)\ , \no
{\cal M}^2(X) &\equiv&
\Phi^2(X)  + H^2(X)   - \tau_3 \not\!\partial \Phi(X)
 - \tau_1 \not\!\partial H(X)\ ,\la{quad1}
\ea
acting on a $n$-dimensional flat Euclidean space (eventually $n = 5$).
Our aim is to evaluate the matrix element of the distribution of the states
operator: namely,
\be
\langle X|\vartheta(Q^2 - D^\dagger D)|Y\rangle =
\int_{c-i\infty}^{c+i\infty}\frac{d t}{2\pi i}\ \frac{\exp\{tQ^2\}}{t}\,
\langle X|\exp\{-tD^\dagger D\}|Y\rangle\ ,\quad c>0\ .\la{a1}
\ee
As a matter of fact, if the positive operator $D^\dagger D$ 
is also of the trace class,
then the trace of $\vartheta(Q^2 - D^\dagger D)$ does represent the number
of the eigenstates of $D^\dagger D$ up to the momentum square $Q^2$.
It is convenient to write the heat kernel in the form
\be
\langle X|\exp\{-tD^\dagger D\}|Y\rangle\equiv
(4\pi t)^{-n/2}\exp\left\{-\frac{(X-Y)^2}{4t}\right\}
\Omega(t|X,Y)\ ,\la{a2}
\ee
where $\Omega(t|X,Y)$ is the so called transport function
which fulfills
\ba
&&\left(\partial_t+\frac{X\cdot\partial}{t}+
D^\dagger_X D_X\right)\Omega(t|X,Y)=0\ ,\\
&&\lim_{t\downarrow 0}\Omega(t|X,Y)={\bf 1}\ ,\la{a3}
\ea
in such a way that
\be
\lim_{t\downarrow 0}\ \langle X|\exp\{-tD^\dagger D\}|Y\rangle=
{\bf 1}\delta^{(n)}(X-Y)\ .\la{a4}
\ee
If we insert Eq.\,\gl{a2} in Eq.\,\gl{a1} and change the integration variable
we obtain
\ba
\langle X|\vartheta(Q^2 - D^\dagger D)|Y\rangle&=&
2Q^n (4\pi)^{-1-n/2}\int_{c-i\infty}^{c+i\infty}
dt\ t^{-1-n/2}\no
&\times&\exp\left\{t-\frac{Q^2(X-Y)^2}{4t}-\frac{i\pi}{2}\right\}
\Omega\left(\frac{t}{Q^2}\left|X,Y\right.\right)\ .\la{a5}
\ea
Now, if we write
\be
\Omega\left(\frac{t}{Q^2}\left|X,Y\right.\right)=
\sum_{k=0}^{[n/2]}t^k a_k(X,Y)Q^{-2k}
+{\cal R}_{[n/2]+1}\left(\frac{t}{Q^2}\left|X,Y\right.\right)\ ,\la{a6}
\ee
it turns out that, by insertion of Eq.\,\gl{a6} into the integral
\gl{a5}, the last term in
the RHS of the above expression becomes sub-leading and negligible
in the limit of very large $Q$.
As a consequence, the leading asymptotic behavior of the matrix
element \gl{a1} in the large $Q$ limit reads \cite{GR}
\ba
\langle X|\vartheta(Q^2 - D^\dagger D)|Y\rangle
&\stackrel{Q\rightarrow\infty}{\sim}&
(4\pi)^{-n/2}\sum_{k=0}^{[n/2]}a_k(X,Y) Q^{n-2k}\no
&\times&\int_{c-i\infty}^{c+i\infty}
\frac{dt}{2\pi i}\ t^{k-1-n/2}\,\exp\left\{t-\frac{Q^2(X-Y)^2}{4t}\right\}\no
&\stackrel{X\rightarrow Y}{\sim}&
(4\pi)^{-n/2}\sum_{k=0}^{[n/2]}a_k(X,Y)\,
\frac{Q^{n-2k}}{\Gamma(1-k+n/2)}\ .\la{a7}
\ea
In the diagonal limit $X=Y$, for $n=4,5$ the relevant
coefficients of the heat kernel
asymptotic expansion take the form \cite{CGZ}
\ba
a_0(X,X)&\equiv& {\bf 1}\ ;\qquad
a_1(X,X)=-{\cal M}^2(X)\ ;\no
a_2(X,X)&=&\frac{1}{2}[{\cal M}^2(X)]^2
-\frac{1}{6}\partial^2{\cal M}^2(X)\ ,\la{a8}
\ea
in such a way that we can finally write the dominant diagonal matrix element --
leading eventually to the five-dimensional Euclidean
effective Lagrange density -- as
\ba
&&\langle X|\vartheta(Q^2 - D^\dagger D)|X\rangle
\stackrel{Q\rightarrow\infty}{\sim}\no
&&\frac{Q^5}{60\pi^3}\left\{
{\bf 1}-\frac{5}{2}Q^{-2}{\cal M}^2(X)+
\frac{15}{8}Q^{-4}[{\cal M}^2(X)]^2
-\frac{5}{8}Q^{-4}\partial^2{\cal M}^2(X)\right\}\ .
\ea

%%%%%%%%%%%%%%%%%%%%%%%%%%%%%%%%%%%%%%%%%%%%%%%%%%%%%%%%%%%%%%%%%%%%%%%%%%%%%%%%%%%

\section{Spectral resolution for Schr\"odinger operators}

In this Appendix we calculate the spectra and
eigenfunctions of the Schr\"odinger operators ${\rm M}_{J}^{11}$ and
${\rm M}_{J}^{22}$
arising in Eq.s\ \gl{massII1} and \gl{massII2}
and presented in the factorized form in Eq.s\
\gl{factor1} and \gl{factor2}. 
Let us choose the inverse mass units, $z = y/M$ 
and introduce the dimensionless operators
\ba
{\rm M}_{J}^{11} &\equiv& M^2 d^+_\Phi \equiv M^2 q^+_\Phi\,q^-_\Phi\ ,
\label{appb1}\\
{\rm M}_{J}^{22} - M^2 +2\Delta_2 &\equiv& M^2 d^+_H \equiv 
M^2 q^+_H\,q^-_H\ ,\label{appb2}\\
q^\pm_\Phi &=& \mp \partial_y + 2 \mbox{\rm tanh}(y)\ ,\label{appb3}\\
q^\pm_H &=& \mp \partial_y + \mbox{\rm tanh}(y)\ .\label{appb4}
\ea
The two operators $d^+_\Phi$ and $d^+_H$ can be embedded
into the supersymmetric ladder
when one adds their supersymmetric partners
\be
d^-_\Phi \equiv q^-_\Phi\,q^+_\Phi  = d^+_H +3\ ,\qquad
d^-_H \equiv q^-_H\,q^+_H = -\partial^2_y +1\ ,
\ee
the latter one describing a free particle propagation. The intertwining
Darboux relations read
\ba
&&q^\pm_\Phi\,d^\mp_\Phi= d^\pm_\Phi\,q^\pm_\Phi\ ,\no
&&q^\pm_H\,d^\mp_H = d^\pm_H\,q^\pm_H\ ,\no
&& q^-_H\,q^-_\Phi\,d^+_\Phi= q^-_H (d^+_H +3) q^-_\Phi =
(-\partial^2_y +4) q^-_H\,q^-_\Phi\ ,\no
&&d^+_\Phi\,q^+_\Phi\,q^+_H = q^+_\Phi\,q^+_H (-\partial^2_y +4)\ ,\no
&&q^-_H\,d^+_H = (-\partial^2_y +1) q^-_H\ ,\no
&&d^+_H\,q^+_H =  q^+_H (-\partial^2_y +1)\ .\la{ladder}
\ea
From these relations one concludes that the
spectra of the operators $d^+_\Phi$ and $d^+_H$ are almost equivalent 
one to each other and, in turn, both equivalent to the
free particle continuous spectrum -- up to a shift of the origin.
In other words,  they consist of the continuous part and of the
zero-modes of the intertwining operators $q^-_H q^-_\Phi$ and
$q^-_H$ respectively.
Owing to  Eq.s\ \gl{ladder} 
their eigenvalues and normalized improper eigenfunctions
are related as follows: namely,
\ba
&&\psi_\Phi(ky) \doteq \frac{q^+_\Phi\,\psi_H(ky)}{\sqrt{k^2 + 4}}  
\doteq
\frac{q^+_\Phi q^+_H\,\psi_0 (ky)}{\sqrt{(k^2 + 1)(k^2 + 4)}}\ ,
\qquad E^c_\Phi(k) = k^2 + 4\ ,\no
&&\psi_H(ky) \doteq
\frac{q^+_H\,\psi_0 (ky)}{\sqrt{k^2 + 1}}\ ,\qquad\quad
E^c_H(k) = k^2 + 1\ ,\no
&& \psi_0 (ky)=\frac{1}{\sqrt{2\pi}} \exp(i k y)\ ,
\qquad \qquad k\in{\bf R}\ . \la{wavfu}
\ea
The imprecise equality $\doteq$ 
signifies a possible discrepancy due to the existence
of zero-modes.
In particular, for the operator $d^+_H$ the continuum starts
from $E^c_H(0)= 1$ and the only zero-mode
$h_0(y) = \left(1/\sqrt{2}\right)\,\mbox{\rm sech}(y)$ just
coincides with the zero-mode of $q^-_H$.
Concerning the operator $d^+_\Phi$, one can easily verify
that  the continuum starts from
$E^c_\Phi(0)= 4$, whereas the
zero-mode
$\phi_0(y) = \left(\sqrt{3}/2\right)\,\mbox{\rm sech}^2(y)$ coincides with
the zero-mode of $q^-_\Phi$.
Furthermore, the second parity-odd proper eigenstate
\be
\phi_1(y) = \left(1/\sqrt{3}\right)\,q^+_\Phi h_0(y) = \sqrt{3/2}\,
{\mbox{\rm tanh}(y)}\,{\mbox{\rm sech}(y)}\ ,
\ee
does coincide with the zero-mode of $q^-_H\,q^-_\Phi$ and
corresponds to the first non-vanishing, discrete
eigenvalue $E^+_\Phi=3$.

%%%%%%%%%%%%%%%%%%%%%%%%%%%%%%%%%%%%%%%%%%%%%%%%%%%%%%%%%%%%%%%%%%%%%%%%%%%%%%%%%%%%%%

\section{Perturbation theory for the first excited state}
In this Appendix we develop the perturbation theory for the first excited
eigenstate and the corresponding eigenvalue of the operator ${\bf M}_{K}$
defined in Eq.~\gl{pertu}. As in the Appendix B
the inverse mass units $z = y/\beta$ are chosen,
so that the dimensionless operator $\Xi \equiv {\bf M}_{K}/ \beta^2$ has
the following components in terms of the operators in Eq.s\,
\gl{appb1}-\gl{appb4}: namely,
\begin{eqnarray}
\Xi = \left\lgroup\begin{array}{cc}
 d^+_\Phi & 0\\
0  & d^+_H
\end{array}\right\rgroup\ + 4\epsilon\, {\mbox{\rm sech}^2(y)}\left\lgroup\begin{array}{cc}
\epsilon\,{\mbox{\rm sinh}^2(y)}
 & \sqrt{1 + \epsilon^2}\
{\mbox{\rm sinh}(y)}\\
\sqrt{1 + \epsilon^2}\
{\mbox{\rm sinh}(y)}
& \epsilon
\end{array}\right\rgroup\ . \la{xiop}
\end{eqnarray}
The second, perturbation term contains the elements of different order in
$\epsilon$. This is the ultimate reason why, in order to compute the 
corrections to the eigenvalues, we have to take into account not only the
eigenfunction of zeroth-order, but also the first-order correction
turns out to be necessary. 
Thus we use the eigenfunction $h_K(y) =[h^-_K(y),h^+_K(y)]$
from Eq.\,\gl{state22},
taking into account that, to the lowest order, the parity-odd function
$h^-_K(y)$ is of order $\epsilon$ and the parity-even one 
is $h^+_K(y) \simeq
h_0(y)= \left(1/\sqrt{2}\right)\,\mbox{\rm sech}(y)$.
To the first order in $\epsilon$,
the function $h^+_K(y)$ is derived from the following equation
\ba
d^+_\Phi\,h^-_K(y) &=&
- 4\epsilon\,{\mbox{\rm tanh}(y)}\,{\mbox{\rm sech}(y)}\,h_0 (y) =
- \frac{4\epsilon}{\sqrt{2}}\,{\mbox{\rm tanh}(y)}\,{\mbox{\rm 
sech}^2(y)}\no
 &=& -\ \frac{\epsilon}{\sqrt{2}}\,q^+_\Phi\,
{\mbox{\rm sech}^2(y)}\ , \la{hminus}
\ea
which follows from the basic definition $\Xi h_K = E_h  h_K $ and
the estimation of $E_h \sim \epsilon^2$. Since there are no normalizable
zero-modes for the operator $q^+_\Phi$ one concludes that
\be
q^-_\Phi\,h^-_K(y) = -\frac{\epsilon}{\sqrt{2}}\,{\mbox{\rm sech}^2(y)}\ ,
\ee 
which represents a first-order differential equation with the normalizable
solution
\be
h^-_K(y) = -  \frac{\epsilon}{\sqrt{2}}\,y\,{\mbox{\rm sech}^2(y)}\ .
\ee
Now, the eigenvalue  $E_h$ can be obtained from its integral representation
 $E_h = \langle h_K\left|\,\Xi\,\right| h_K\rangle$ in terms of
the normalizable function $\langle h_K| h_K\rangle = 1$. To the first order in
$\epsilon$ it reads eventually
\be
E_h  = 4 \int^{+\infty}_{-\infty}dy\ h_0 (y) \left\{\epsilon^2
{\mbox{\rm sech}^2(y)}\,h_0 (y) + \epsilon\,{\mbox{\rm tanh}(y)}\,
{\mbox{\rm sech}(y)}\,h^-_K(y)\right\} = 2 \epsilon^2\ .
\ee

%%%%%%%%%%%%%%%%%%%%%%%%%%%%%%%%%%%%%%%%%%%%%%%%%%%%%%%%%%%%%%%%%%%%%%%%%%%%%%%%%%%%%%%%%%

\section{Perturbation theory in the presence of defects}

The inclusion of a small regular defect is described by the scalar functions
${\rm f}_\Phi (z), {\rm f}_H (z)$ and leads to the change of the 
mass matrix ${\bf M}_{K}$ presented in Eq.~\gl{defmA} for the ansatz $(A)$ 
and  in Eq.~\gl{defmB} for the ansatz $(B)$.
If we introduce the dimensionless operators $\Xi_j\, ,\quad j = A,B\,$, then
according to the notations of Eq.~\gl{xiop} we obtain
\begin{eqnarray}
&& \Xi_j = \Xi_K + {\bf \Delta}_j\ , \la{xif}\\
&&\Delta_A^{11} \approx
 \epsilon^2\,{\mbox{\rm sech}^2(y)}\,
[\,4 \xi
 + 2 \xi^2 - 12 y\ {\mbox{\rm tanh}(y)}\,]\ ,\no
&&\Delta_A^{12} = \Delta_A^{21} 
\approx  4 \epsilon\,\xi\ {\mbox{\rm sech}^2(y)}\,
{\mbox{\rm sinh}(y)}\ , \\
&& \Delta_A^{22} 
\approx \epsilon^2\, {\mbox{\rm sech}^2(y)}\, [\,12 \xi
 + 6 \xi^2 - 4 y\ {\mbox{\rm tanh}(y)}\,]\ ,\no
&& \Delta_B^{11} \approx \epsilon^2\,{\mbox{\rm sech}^2(y)} 
\left[\,4 \xi\  + 2 \xi^2
+ 3 (\gamma - 4 \xi\  - 2 \xi^2)\,
y\,{\mbox{\rm tanh} (y)}+ 
3 \gamma {\mbox{\rm sinh}^2 ( y)}\,\right]\ ,\no
&& \Delta_B^{12} \approx \Delta_B^{21} =  4 \xi  
\epsilon\,{\mbox{\rm sech}^2 (y)}
{\mbox{\rm sinh}( y)}\ , \\
&&\Delta_B^{22} \approx \epsilon^2\,{\mbox{\rm sech}^2(y)} 
\left[\,12 \xi\  + 6 \xi^2
+  (\gamma - 4 \xi\  - 2 \xi^2)\,
y \,{\mbox{\rm tanh} (y)} 
+  \gamma {\mbox{\rm sinh}^2 (y)}\,\right]\  ,\nonumber
\end{eqnarray}
One can see that again the perturbation matrices contain elements of different
order in $\epsilon$ and therefore the calculation of energy levels needs
first to derive the perturbed eigenfunctions to a relevant order in $\epsilon$
as it has been done in the Appendix C.

To the relevant order in $\epsilon$, the eigenvalue equations
for the scalar wave function components $s_A^+(y), s_A^-(y)$ 
follow from the matrix elements of the operator in Eq.~\gl{xif},
\ba
&&(d^+_\Phi - E_s)\ s_j^+(y) +  \Bigl[\,4 \epsilon^2\,
{\mbox{\rm tanh}^2(y)} 
+ \Delta_j^{11}\,\Bigr]\  s_j^+(y)\no
&& +
 4\epsilon\ (1 +\xi )\ {\mbox{\rm sech}^2(y)}\ 
 {\mbox{\rm sinh}(y)}\ s_j^-(y) = 0\ ,
\no
&&(d^+_H - E_s)\ s_j^-(y) +  \Bigl[\,4 \epsilon^2\,{\mbox{\rm sech}^2(y)}\,
 + \Delta_j^{22}\,\Bigr]\ s_j^-(y)\no
&& +
 4\epsilon\ (1 +\xi )\ {\mbox{\rm sech}^2(y)}\ 
 {\mbox{\rm sinh}(y)}\ s_j^+(y) = 0\ .
\la{eqmass} 
\ea 
These equations have terms of different order in  $\epsilon$ in two sectors of
solutions. 

Actually, for even-odd solutions with  $s_j^\pm (y) = \phi^\pm_j (y)$,
the upper even component
$\phi_j^+ = \phi^{+ (0)} + \epsilon^2 \phi_j^{+ (1)} + {\cal O}(\epsilon^4)$
 contains the zero mode \gl{state1} of the operator 
$d^+_\Phi$ as a leading term. Meanwhile, the lower odd component $\phi_j^-$ has
an order of $\epsilon$ and coincides with the unperturbed one -- 
see Eq.~\gl{state1} in its functional form. 
Indeed, as one expects that the lightest eigenvalues are of the order
$E_s \sim  \epsilon^2$, then the
dominating part of the second Eq.~\gl{eqmass} reads,
\be
d^+_H\, \phi_j^{- (0)} +
 4\epsilon\ (1 +\xi )\ {\mbox{\rm tanh}(y)}\,{\mbox{\rm sech}(y)} 
\, \phi^{+ (0)} = 0.
\ee
Comparing this equation with the unperturbed one, corresponding to
$\xi = 0$, from Eq.~\gl{state1} one immediately finds  that
\be
 \phi_j^{- (0)}(y)  = - \sqrt{\frac{3\beta}{4}}\,\epsilon\ (1 +\xi )\
{\mbox{\rm sinh} (y)}\,{\mbox{\rm sech}^2 (y)} \equiv
\phi^{- (0)}(y)\ .
\ee 
On the other hand, as $d^+_\Phi\,\phi^{+ (0)}=0$
the first Eq.~\gl{eqmass} determines the next-to-leading contribution  
$\phi_j^{+ (1)}$
into the even part of the wave function.
Now this equation allows us to obtain both the 
approximate eigenvalue and the correction term $\phi_j^{+ (1)}$.
After projection onto the zero-mode $\phi_j^{+ (0)}$ of the operator
$d^+_\Phi$, one gets rid of the unknown function  $\phi_j^{+ (1)}$ and
calculates eventually the branon mass: namely,
\ba
\left(m^j_\phi\right)^2 &=&  
\beta^2 E_\phi = \beta^2 \int^{+\infty}_{-\infty} dy \
\left\{\Bigl[\,4 \epsilon^2\,
{\mbox{\rm tanh}^2(y)} 
+ \Delta_j^{11}\,\Bigr]\left[\,
\phi^{+ (0)}(y)\,\right]^2 \right.\no
&&\left.+
4 \epsilon\, (1 +\xi ) \ {\mbox{\rm sech}^2(y)}\ {\mbox{\rm sinh}(y)}\,
\phi^{- (0)}(y)\,\phi^{+ (0)}(y) \right\}\ .
\ea
In the sector of odd-even solutions  $s_j^\pm (y) = h^\mp_j (y)$
the lightest-state wave function has the opposite order in  $\epsilon$, 
{\it i.e.},
$$ h^{- (0)}_j(y) \sim \epsilon,\ \ h^+_j (y) = h^{+ (0)} + 
\epsilon^2 h_j^{+ (1)} + {\cal O}(\epsilon^4)\ .$$ 
Thus the role of two Eq.s~\gl{eqmass} 
is merely interchanged. In particular, it can be easily derived that
\ba
&& h^{- (0)}_j(y) = -\ \epsilon\,
(1 +\xi )\ y\,{\mbox{\rm sech}^2(y)}/\sqrt{2} 
\equiv  h^{- (0)}(y)\ ,\no
&&h^{+ (0)}(y) =\ \mbox{\rm sech}(y)/\sqrt{2}\ ,
\ea
to be compared with Eq.~\gl{state22}.
Respectively, when we project onto the zero-mode $h^{+ (0)}_j(y)$
of the operator $d^+_H$,
the second equation helps us to obtain the Higgs boson mass,
\ba
\left(m^j_h\right)^2 &=& \beta^2  E_h = \beta^2 \int^{+\infty}_{-\infty} dy \
\left\{\Bigl[\,4 \epsilon^2\,{\mbox{\rm sech}^2(y)}\,
 + \Delta_j^{22}\,\Bigr]\left[\,h^{+ (0)}(y)\,\right]^2
\right.\no
&&\left.
+ 4\epsilon\,
(1 +\xi )\ {\mbox{\rm sech}^2(y)}\,{\mbox{\rm sinh}(y)}\,  
h^{- (0)}(y)\,h^{+ (0)}(y) \right\}\ .
\ea

\end{document}